\documentclass{aa}  

\usepackage{graphicx}
\usepackage{txfonts}
\usepackage{lipsum}
\usepackage{lscape}             

\usepackage{bm}

\usepackage{placeins}           

\usepackage{amsmath}
\usepackage{xcolor}
\usepackage[colorlinks=true,linkcolor=blue,citecolor=blue]{hyperref}
\usepackage{ulem}

\begin{document}

   \title{Acoustic instability at shock-wave precursors}



   \author{A. Capanema\inst{1}\fnmsep\inst{2}\fnmsep\thanks{Corresponding author: \texttt{antonio.capanema@gssi.it}}  
   \and P. Blasi\inst{1}\fnmsep\inst{2}
   \and E. Sobacchi\inst{1}\fnmsep\inst{2}
        }

   \institute{Gran Sasso Science Institute (GSSI), L'Aquila, Italy \and INFN -- Laboratori Nazionali del Gran Sasso (LNGS), L'Aquila, Italy}

   \date{Received September 30, 2025}

  \abstract
  {Magnetic field amplification is an integral part of the process of particle acceleration at non-relativistic shocks. It is necessary to reach the maximum energies required by observations, especially in supernova remnants, thought to be sources of the bulk of Galactic cosmic rays. Such amplification can be caused by the acoustic instability that develops when small density perturbations interact with the cosmic-ray pressure gradient in the upstream of a cosmic-ray-modified shock. The vorticity induced by the nonlinear development of the instability may lead to turbulence, which amplifies the pre-existing magnetic fields. To study this phenomenon, we use the PLUTO code to carry out 2D (and some 3D) magnetohydrodynamical simulations of the evolution of small density perturbations in the presence of an assigned cosmic-ray pressure gradient. Adopting more realistic values of Mach number and cosmic-ray acceleration efficiency than previously assumed in the literature, we show that the acoustic instability can transform small density perturbations into large nonlinear structures while the fluid crosses the precursor region of a cosmic-ray-modified shock. We study the power spectrum of turbulent magnetic fluctuations that may be important to scatter particles. We comment on the possible constructive interference between acoustic and non-resonant streaming instabilities. We discuss limitations of previous and current numerical investigations in accessing spatial scales where turbulence is expected to turn nonlinear, and outline perspectives for future investigations.}


   \maketitle
   \nolinenumbers

\section{Introduction}
\label{sec:intro}

The influence of cosmic rays (CRs) over their own transport is crucial to explain numerous phenomena: from the diffusive properties of astrophysical plasmas, to the maximum energy reached by accelerators, to the subsequent spectra we measure at the Earth (see \citet[]{Blasi2019} for a review of these effects). In recent decades, substantial progress has been made in understanding how CRs shape the environment in which they are produced. This is especially true for the vicinity of shock waves, which are expected to be key sites of CR acceleration across a wide range of astrophysical settings, from supernova remnants (SNRs) \cite[]{Blasi2013,Amato2014} to galaxy clusters \cite[]{Review2007}.

Despite the success of early theoretical approaches to diffusive shock acceleration (DSA), in which CRs were mostly treated as test particles \citep{1977DoSSR.234.1306K,Bell:1978dv}, it soon became clear that to account for CRs up to the knee region or even higher energy CRs, efficient shock acceleration and strong magnetic fields, likely powered by CR-driven instabilities, were needed. In particular, the resonant streaming instability was invoked as a process responsible for magnetic field amplification (MFA) upstream of non-relativistic shocks \citep{Skilling:1975kf,Bell:1978dv} and proven to lead to acceleration up to tens of TeV in typical SNRs \citep{Lagage:1983zz}. This process has the advantage of producing Alfv\'en waves at the right scale to scatter the same particles responsible for exciting the instability. On the other hand, the saturation of the instability to $\delta B/B\lesssim 1$ implies that this process cannot account for particle acceleration to the knee. More recently, a non-resonant branch of the streaming instability \citep{Bell:2004hhd} received much attention: this instability is excited by the current induced by the CR particles escaping the upstream region of a shock and can lead to $\delta B/B\gg 1$ due to its non-resonant nature. The instability is expected to saturate when particles start scattering on the self-generated perturbations. The implications of the excitation of the non-resonant modes in terms of maximum energy of particles accelerated in SNR shocks have been discussed in a number of articles \cite[]{Bell2013,Schure2013,Cristofari2020,Cristofari:2021hbc}. The consensus among these works is that although the instability appears to be very effective in amplifying the magnetic field on the relevant scales for very fast shocks in dense environments, the maximum energy at the beginning of the Sedov phase of SNRs typically falls short of PeV by at least one order of magnitude. This leaves open the possibility that other instabilities may lead to more effective MFA and perhaps more effective particle acceleration.

One such mechanism is associated with the pressure gradient that unavoidably forms upstream of a shock due to energy-dependent particle transport, which may induce the so-called Drury or acoustic instability (AI) in the presence of density inhomogeneities. This instability was first proposed by \citet{Drury:1984do} and studied in detail by \citet{Drury:1986vd} (hereafter, DF) using a two-fluid prescription in the absence of magnetic fields. This instability can overcome dissipation by CR diffusion \citep{Ptuskin:1981br} if the scale height of the CR pressure is smaller than the ratio between the local diffusion coefficient and sound speed, a condition naturally met at shock wave precursors. This results in an exponential growth of small inhomogeneities present in the interstellar medium (ISM) as they approach the shock surface. Numerical solutions to the equations in DF confirm the onset of the instability and the growth of density perturbations to nonlinear levels, if enough time is granted.

These findings have prompted further investigation of AI, especially in the context of CR-modified shocks (see \citet{Zank:1990nx, Kang:1992vw, Ryu:1993nk, Begelman:1994bo, 1999JPlPh..61..553W} and references therein), although the nonlinear evolution of the system and the implications for MFA remained poorly understood. \citet{Beresnyak:2009pi} proposed that the CR pressure gradient acting on large-amplitude density inhomogeneities upstream of a strong shock could induce turbulence and that this might reflect in higher maximum energies of the accelerated particles. In this picture, the solenoidal fluid motions act as a small-scale dynamo, amplifying magnetic fields by at least one order of magnitude via the stretch-twist-fold mechanism. 

\citet{Drury:2012xb} (hereafter, DD) and \citet{Downes:2014rta} ran 2D magnetohydrodynamical (MHD) simulations of the precursor region, assuming that CRs were only responsible for the formation of an assigned pressure gradient. These simulations confirmed that MFA actually occurs. In 2D simulations, slightly more effective magnetic energy amplification is found, with a weak predominance of power at larger scales compared to the results of 3D simulations. More recently, \citet{delValle:2016zsp} (hereafter, VLS) explored and expanded on this idea by means of 3D MHD simulations of SNR shock precursors. Their work provided quantitative insights on the development of small-scale dynamo, turbulence, and scales where equipartition between fluid and magnetic turbulence should be expected. We will comment further on these results throughout this paper.

Much of this previous work is based on a few critical assumptions: the first is that a very large part (typically 60\%) of the ram pressure is channeled into accelerated particles. Although early work on the nonlinear theory of DSA appeared to suggest that such an energy conversion efficiency would be feasible \cite[]{Malkov2001,Blasi2002}, later work, based on nonlinear DSA models \citep{Morlino:2011di} and simulations \citep{Caprioli:2013dca}, suggests a typical efficiency $\lesssim 10\%$. The question of MFA in higher Mach numbers with more realistic shock precursors remains open. The second assumption is that the pre-existing density fluctuations in the upstream plasma are already relatively large: this leaves little room for AI and in fact much of the previous work on the topic focuses on the possibility that the CR pressure gradient may induce turbulence in an inhomogeneous plasma. Both of these assumptions were adopted in order to minimize the demands on the numerical machinery to be employed to simulate the instability.

Here, we focus on three main goals. 1) We provide an analytic framework to assess the importance of AI in the precursor in the presence of a magnetic field, and confirm the linear evolution of the instability with 2D MHD simulations. 2) We bridge the gap with previous work, by starting with small density perturbations and investigating the possibility that turbulence may be formed due to AI first and warping of the magnetic field lines later. 3) We use more realistic CR acceleration efficiencies and Mach numbers that are more appropriate to SNR in the beginning of the Sedov phase ($M\sim$~few hundreds). Special care is dedicated to discussing the scales accessible to the simulations (namely with negligible numerical damping), in 2D and 3D, compared with the wavenumbers that are excited by AI.

This article is organized as follows. In Sect.~\ref{sec:instability}, we present an analytic derivation of the dispersion relation and the growth rate of AI. We derive simple expressions for the maximum number of e-folds by which small upstream perturbations can grow before reaching the shock. In Sect.~\ref{sec:sims}, we use a setup similar to that of DD to perform MHD simulations of the instability, comparing the observed growth rates with the expected ones obtained analytically. We also study the nonlinear stage of the system, discuss MFA, and calculate the power spectrum of magnetic perturbations. Finally, in Sect.~\ref{sec:powerspec} we comment on the comparison between MFA due to the CR pressure gradient and the Bell instability, speculating about a possible cooperation between the two processes. We summarize our findings and present our conclusions in Sect.~\ref{sec:concl}.

\section{Acoustic instability}
\label{sec:instability}

We consider a plasma with density $\rho$, velocity $\mathbf{u}$, pressure $P$, adiabatic index $\gamma$, and magnetic field $\mathbf{B}$. Since we aim to describe the plasma dynamics in the upstream of a non-relativistic shock, where a CR pressure gradient is expected, the MHD equations should include a force term due to the presence of a CR pressure gradient $\nabla P_{\rm CR}$:
\begin{align}
    &\frac{\partial \rho}{\partial t} + \nabla\cdot(\rho\mathbf{u})=0~,\label{eq:MHD1}\\
    &\frac{\partial \mathbf{u}}{\partial t}+(\mathbf{u\cdot\nabla})\mathbf{u} = -\frac{\nabla P}{\rho} - \frac{\nabla P_{\rm CR}}{\rho} + \frac{(\nabla \times \mathbf{B})\times \mathbf{B}}{4\pi \rho}~,\label{eq:MHD2}\\
    &\frac{\partial}{\partial t}\left(\frac{P}{\rho^\gamma}\right) + (\mathbf{u}\cdot \nabla)\left(\frac{P}{\rho^\gamma}\right)=0~,\label{eq:MHD3}\\
    &\frac{\partial \mathbf{B}}{\partial t} = \nabla\times(\mathbf{u}\times\mathbf{B})~,\label{eq:MHD4}\\
    &\nabla\cdot\mathbf{B}=0\label{eq:MHD5}\;.
\end{align}

DF modeled the evolution of the CR pressure through the CR transport equation with diffusion coefficient $D$, and did not consider the effect of the magnetic field. By simultaneously perturbing $\rho$, $\mathbf{u}$, $P$, and $P_{\rm CR}$, using a rigorous perturbative approach, they found that these perturbations grow when the length scale of the CR pressure gradient, $P_{\rm CR}/\nabla P_{\rm CR}$, is smaller than $D/c_{\rm s}$, where $c_{\rm s}=\sqrt{\gamma P_0/\rho_0}$ is the sound speed. If this condition is not met, the perturbations are damped due to friction with the CRs \citep{Ptuskin:1981br}. In shock precursors, damping effects can be safely neglected because $P_{\rm CR}/\nabla P_{\rm CR} \sim D/v_{\rm sh}\ll D/c_{\rm s}$, where $v_{\rm sh}\gg c_{\rm s}$ is the shock speed. As such, for the case of a shock precursor, it is not necessary to go through the elegant but cumbersome calculations of DF, and the underlying physics describing the instability can be captured by perturbing only the plasma quantities $\rho$, $\mathbf{u}$, and $P$, while assuming that $\nabla P_{\rm CR}$ is constant. Although this neglects the back-reaction of the instability onto CRs, it also reduces the complexity of the problem to a purely MHD one. We will comment later on the implications of this assumption and on the reaction of the system to the enhanced scattering that may follow the excitation of the instability.

A clear advantage of the MHD formulation is that the dispersion relation and the growth rate of the instability can be derived straightforwardly. In the reference frame of the plasma, we can assume that the background plasma is uniform\footnote{Strictly speaking, one must introduce a balancing force $\nabla P_{\rm CR}/\rho_0$ in order for this to be a solution of Eq.~\eqref{eq:MHD2}. Clearly, this term does not affect the perturbed equations.} for Eqs.~\eqref{eq:MHD1}-\eqref{eq:MHD5}, with $\rho=\rho_0$, $\mathbf{u}=0$, $P=P_0$, and $\mathbf{B}=\mathbf{B}_0$ while we introduce small perturbations $\delta \rho$, $\delta\mathbf{u}$, $\delta P$, and $\delta\mathbf{B}$ into the system. This approximation is valid for perturbations on scales much smaller than the scale of the CR pressure gradient. We assume that the perturbations are proportional to $\exp[i(\mathbf{k}\cdot\mathbf{x}-\omega t)]$. The dispersion relation is
\begin{multline}\label{eq:det0}
    (\omega^2 - v_{\rm A}^2k_\parallel^2)\left[\omega^4 - \omega^2\left(k^2c_{\rm ms}^2 + \frac{i\mathbf{k}\cdot\nabla P_{\rm CR}}{\rho_0}\right) \right. \\ \left. + k^2k_\parallel^2v_{\rm A}^2c_{\rm s}^2 + iv_{\rm A}^2k^2k_\parallel \frac{\partial_\parallel P_{\rm CR}}{\rho_0}\right] = 0~,
\end{multline}
where $v_{\rm A}=B_0/\sqrt{4\pi\rho_0}$ is the Alfvén speed, $c_{\rm ms} = \sqrt{c_{\rm s}^2 + v_{\rm A}^2}$ is the magnetosonic speed, and the $\parallel$ subscript refers to components parallel to $\mathbf{B}_0$.

In addition to the usual Alfvén waves ($\omega^2= v_{\rm A}^2 k^2_\parallel$), the dispersion relation shows that there are unstable modes, roots of the expression inside the square brackets. When $k_\parallel = 0$, the dispersion relation is
\begin{equation}\label{eq:kperp-dispersion}
    \omega^2 = c_{\rm ms}^2 k^2 + \frac{i\mathbf{k}\cdot \nabla P_{\rm CR}}{\rho_0}~,
\end{equation}
while for $|\mathbf{k}|=k_\parallel$ it is the same as in the absence of a magnetic field, namely
\begin{equation}\label{eq:kpar-dispersion}
    \omega^2 = c_{\rm s}^2 k^2 + \frac{i\mathbf{k}\cdot \nabla P_{\rm CR}}{\rho_0}~.
\end{equation}
The growth rate of the instability, $\Gamma$, is characterized by two distinct regimes, which depend on the ratio between the real and imaginary parts of $\omega^2$. For example, Eq. (\ref{eq:kpar-dispersion}) with $\mathbf{k}\parallel \nabla P_{\rm CR}$ yields
\begin{equation}\label{xc}
    \Gamma = \begin{cases}
        \sqrt{\dfrac{k|\nabla P_{\rm CR}|}{2\rho_0}}~, & k\ll \dfrac{|\nabla P_{\rm CR}|}{\rho_0 c_{\rm s}^2}
        \\[2.5ex]
        \dfrac{|\nabla P_{\rm CR}|}{2\rho_0 c_{\rm s}}~, & k\gg \dfrac{|\nabla P_{\rm CR}|}{\rho_0 c_{\rm s}^2}
    \end{cases}~.
\end{equation}
The real part of $\omega$, which determines the phase velocity of the perturbations, is equal to $\Gamma$ for long wavelengths ($k\ll |\nabla P_{\rm CR}|/\rho_0 c_{\rm s}^2$). Instead, the real part of $\omega$ is equal to $c_{\rm s} k$ for short wavelengths ($k\gg |\nabla P_{\rm CR}|/\rho_0 c_{\rm s}^2$). In the latter regime, the perturbations are modified sound waves because $\Gamma\ll c_{\rm s} k$. In the other regime, the condition that the perturbations must occur on scales smaller than the size of the precursor, $L\approx P_{\rm CR}/\nabla P_{\rm CR}$, implies that the sonic Mach number must be $M_s=u_0/c_{\rm s}\gg \xi_{\rm CR}^{-1/2}$, where $\xi_{\rm CR}=P_{\rm CR}/\rho_0 u_0^2$. This condition is certainly satisfied for shocks of astrophysical relevance.

To better understand the nature of the instability, let us consider only modes with $\mathbf{k}\parallel\nabla P_{\rm CR}$ and compute the eigenmodes of the system. If the density perturbation is given by $\delta\rho = \delta\rho_0 \,e^{i(\mathbf{k}\cdot \mathbf{x}-\omega t)}$, one can show that pressure perturbations are $\delta P = \delta\rho_0 c_s^2\, e^{i(\mathbf{k}\cdot \mathbf{x}-\omega t)}$, just like in regular sound waves, while velocity perturbations become
\begin{equation}\label{eq:velocity-eigenmode}
    \delta\mathbf{u} = \pm \frac{\Omega}{k}\frac{\delta\rho_0}{\rho_0}\,e^{i(\mathbf{k}\cdot\mathbf{x}-\omega t + \phi/2)}\,\hat{\mathbf{k}}~,
\end{equation}
where $\Omega = |\omega|=\sqrt[4]{c_s^4k^4 + (\mathbf{k}\cdot \nabla P_{\rm CR}/\rho_0)^2}$ and $\phi=\arg(\omega^2)=\arctan(\mathbf{k}\cdot \nabla P_{\rm CR}/\rho_0c_s^2k^2)$ is the principal value ($-\pi/2<\phi\leq\pi/2$). The key point behind the instability is that velocity and pressure perturbations are out of phase, as was also noted in previous work (\textit{e.g.} \citealt{Begelman:1994bo}). Consider only the positive sign of Eq.~(\ref{eq:velocity-eigenmode}). Assuming $\mathbf{k}\cdot\nabla P_{\rm CR}=k|\nabla P_{\rm CR}|$ (that is, $\omega_R,\omega_I>0$), we have $0 < \phi < \pi/2$ and therefore the pressure lags the velocity in phase, leading to instability. This is similar to pushing a swing soon after it starts moving away, doing work to increase its amplitude of oscillation. Meanwhile, we find $-\pi/2<\phi<0$ if we take waves in the opposite direction ($\mathbf{k}\cdot\nabla P_{\rm CR}=-k|\nabla P_{\rm CR}|$ or equivalently $\omega_R,\omega_I<0$), making pressure precede velocity in phase. The result is like pushing a swing too early, damping its motion. The equivalent analysis taking the negative sign for $\delta \mathbf{u}$ in Eq.~(\ref{eq:velocity-eigenmode}) reverses the effects of leading/lagging pressure and is left for the reader to carry out.

\begin{figure}[t!]
    \centering
    \includegraphics[width=0.95\linewidth]{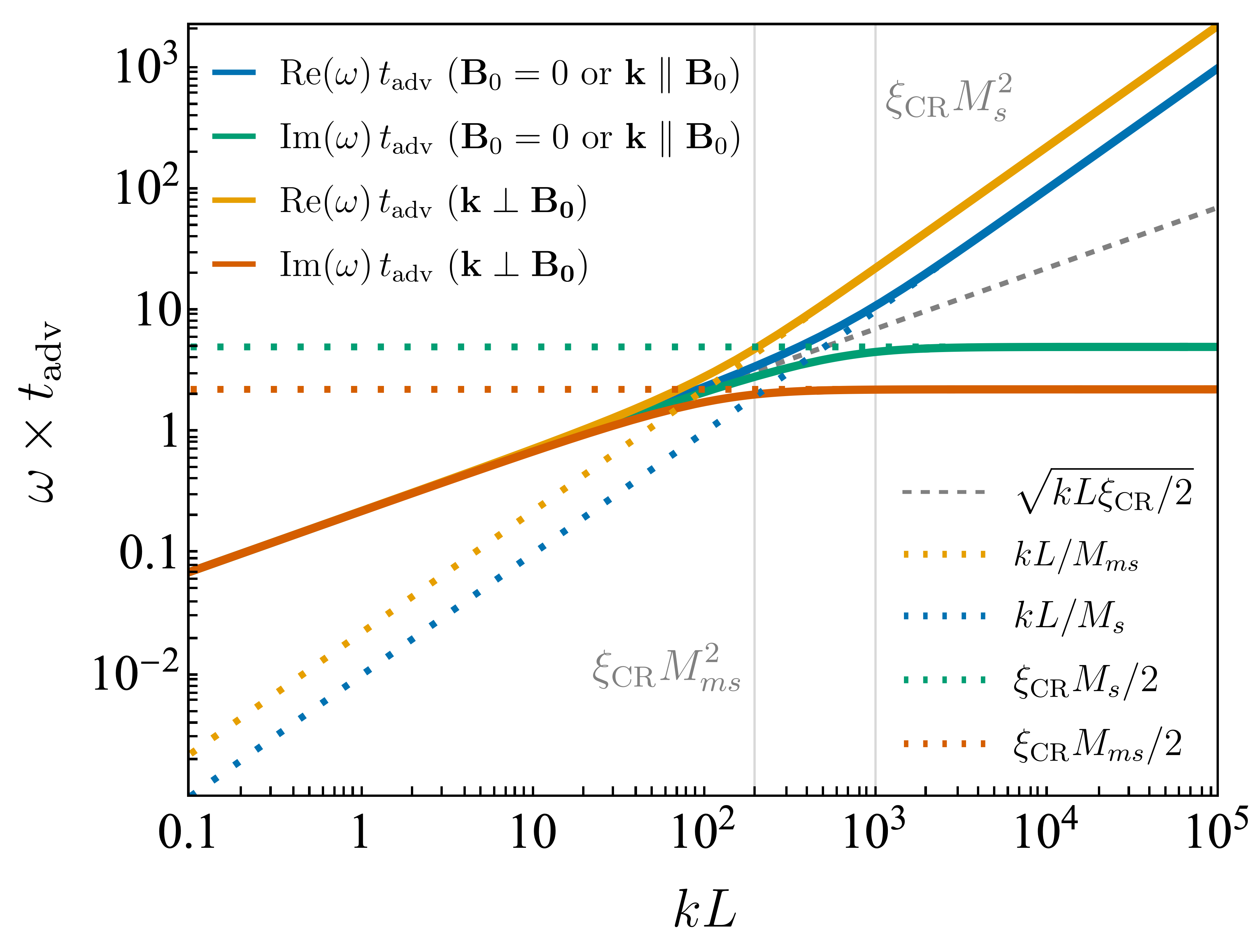}
    \caption{Real and imaginary parts of $\omega$. The CR pressure gradient is constant, $|\nabla P_{\rm CR}|=\xi_{\rm CR}\rho_0u_0^2/L$, and the perturbation wave vector $\mathbf{k}$ is aligned with $\nabla P_{\rm CR}$. We assume $\xi_{\rm CR}=0.1$, $M_s=100$, and $M_{\rm ms}=100/\sqrt{5}$. The dashed and dotted lines show the asymptotic behavior of $\omega$.}
    \label{fig:dispersions}
\end{figure}

Fig.~\ref{fig:dispersions} shows the real and complex parts of $\omega$ as functions of $k$, as predicted by Eqs.~\eqref{eq:kperp-dispersion} and \eqref{eq:kpar-dispersion}, along with their asymptotic behaviors. We assume that $\mathbf{k}\parallel\nabla P_{\rm CR}$ and the CR pressure gradient is constant, $|\nabla P_{\rm CR}|=\xi_{\rm CR}\rho_0u_0^2/L$, where $L$ is the length of the precursor and $\xi_{\rm CR}$ is the fraction of the shock ram pressure converted into CRs. The number of e-folds available for the instability to grow on the precursor scale can be estimated as $\Gamma t_{\rm adv}$, where $t_{\rm adv}\approx L/u_0$ is the advection timescale. In the absence of magnetic fields, Eq.~\eqref{xc} gives
\begin{equation}\label{eq:drury-growth}
    \Gamma t_{\rm adv} = \begin{cases}
        \sqrt{\dfrac{kL\xi_{\rm CR}}{2}}~, & kL\ll \xi_{\rm CR}M_s^2\\[2.5ex]
        \dfrac{\xi_{\rm CR} M_s}{2} ~, & kL\gg \xi_{\rm CR}M_s^2
    \end{cases}~.
\end{equation}
The instability develops when $\Gamma t_{\rm adv}>$~a few. If the magnetic field is perpendicular to $\nabla P_{\rm CR}$ (and consequently to $\mathbf{k}$), $M_s$ should be replaced by the magnetosonic Mach number, $M_{\rm ms}=u_0/c_{\rm ms}$.

\begin{figure*}[t]
    \centering
    \includegraphics[width=18cm]{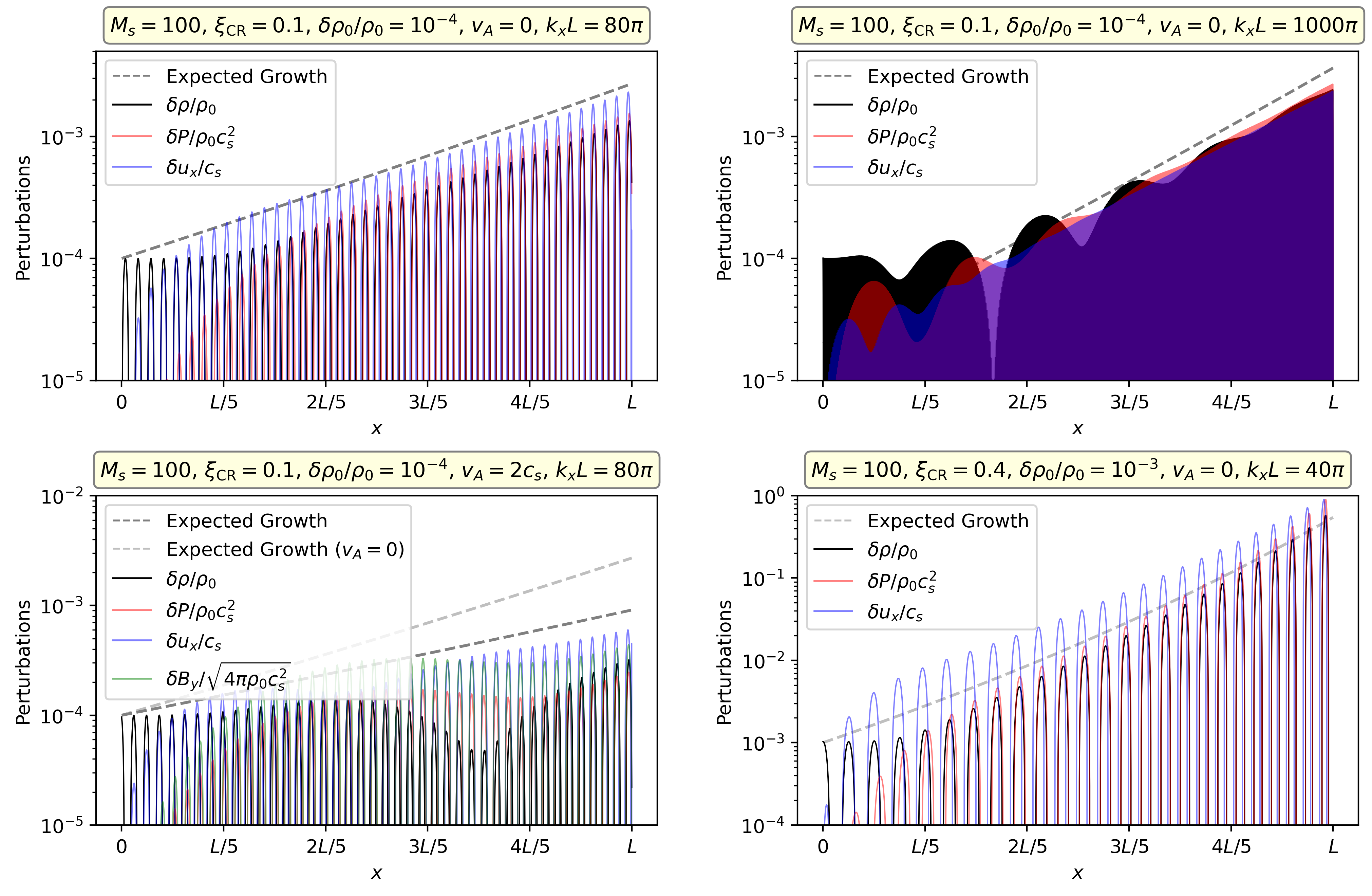}
    \caption{Quasi-stationary $x$-profiles of the dimensionless density, pressure, velocity, and magnetic field perturbations from $k_x$ modes. We show the simulation results as solid lines and the expected growths for AI derived in Sect.~\ref{sec:instability} as dashed lines. In the bottom left panel, the initial magnetic field pointed along $y$, and the lighter gray line corresponds to the expected growth in the absence of magnetic fields.} 
    \label{fig:kx-steady-states}
\end{figure*}

\section{Simulations}
\label{sec:sims}

In order to investigate the linear evolution of AI discussed in Sect.~\ref{sec:instability} and, more importantly, in order to study its nonlinear evolution, we perform MHD simulations using the PLUTO code \citep{Mignone:2007iw}. Although most of our simulations are in 2D, we also compare the results of selected cases of physical interest with 3D simulations. Inspired by DD, our simulation box consists of a uniform 2D grid of dimensions $N_x \times N_y$ embedded in a Cartesian coordinate system with dimensions $ [0,L]\times [0,L/8]$. This region represents the precursor of a shock wave in the shock rest frame, with the shock itself lying just outside the right boundary at $x=L$. The $x=0$ boundary represents upstream infinity (\textit{i.e.} the ISM), where plasma of density $\rho_0$, pressure $P_0$, magnetic field $\mathbf{B}_0$, and adiabatic index $\gamma=5/3$ enters the simulation box at the shock speed $\mathbf{u}_0=u_0\,\hat{\mathbf{x}}$.

The CR pressure gradient is introduced as a body force using the ``\texttt{VECTOR}'' prescription in PLUTO. For simplicity, we assume that the CR pressure increases linearly from zero at $x=0$ to a fraction $\xi_{\rm CR}$ of the shock ram pressure at $x=L$:
\begin{equation}
    P_{\rm CR}(x) = \xi_{\rm CR}\,\rho_0u_0^2\,\frac{x}{L} ~, \qquad \nabla P_{\rm CR}(x) = \xi_{\rm CR}\,\frac{\rho_0u_0^2}{L}\,\hat{\mathbf{x}}~.
\end{equation}
The presence of $\nabla P_{\rm CR}$ induces non-uniform profiles along $x$ in the density, velocity, pressure, and magnetic field of the background plasma. The steady-state profiles can be found by solving Eqs.~\eqref{eq:MHD1}-\eqref{eq:MHD5} in their conservative form: 
\begin{equation}\label{eq:bkg-profiles}
    U(x)\equiv\frac{u(x)}{u_0}=\frac{\rho_0}{\rho(x)}=\left[\frac{P_0}{P(x)}\right]^{1/\gamma}=\frac{B_{y,0}}{B_y(x)}
\end{equation}
and $B_{x}(x)=B_{x,0}$, where $U(x)$ satisfies
\begin{equation}\label{eq:U-withB}
    U + \frac{U^{-\gamma}}{\gamma M_s^2} + \frac{U^{-2}}{2M_{{\rm A},y}^2} + \xi_{\rm CR}\frac{x}{L} = 1 + \frac{1}{\gamma M_s^2} + \frac{1}{2M_{{\rm A},y}^2}~,
\end{equation}
where $M_{{\rm A},y}=u_0/v_{{\rm A},y}$ and $v_{{\rm A},y}=B_{y,0}/\sqrt{4\pi\rho_0}$. Note that this is different from the Alfvénic Mach number $M_A=u_0/v_A$.

Eq.~\eqref{eq:U-withB} can be solved numerically to find $U(x)$ exactly. In the limit of high Mach numbers, one can approximate $U(x)\approx 1-\xi_{\rm CR}x/L$. The resulting density, velocity, pressure, and magnetic field profiles are used as the initial conditions for the simulations. The time it takes for a fluid element entering from the left at $x=0$ to reach a position $x$ inside the box in the presence of a shock precursor is
\begin{equation}\label{eq:time}
    t(x) = \int_0^x\frac{dx'}{u(x')} \approx \frac{L}{u_0} \frac{1}{\xi_{\rm CR}}\ln\left(\frac{1}{1-\xi_{\rm CR}x/L}\right) \gtrsim \frac{x}{u_0}~.
\end{equation}

PLUTO implements boundary conditions (BCs) in each timestep by updating the value of physical quantities in inactive ``ghost'' cells surrounding the active $N_x\times N_y$ simulation grid. We use periodic BCs at the $y=0$ and $y=L/8$ boundaries and an outflow (\textit{i.e.} zero-gradient) BC at the $x=L$ boundary, so that the plasma cannot flow back from the shock into the precursor. We extrapolate the variables into the ghost cells at $x=0$ to avoid discontinuities of the derivatives. To avoid unphysical pressure waves that propagate back into the simulation box, we set the pressure at the right boundary to a very small value.

In Sect.~\ref{sec:linear}, we used the ``\texttt{HLLC}'' Riemann solver with ``\texttt{MP5}'' reconstruction and 3$^{\rm rd}$ order Runge-Kutta time-stepping (with Courant number $\mathtt{CFL}=0.4$ for 2D simulations and $\mathtt{CFL}=0.3$ for 3D simulations), while in Sect.~\ref{sec:nonlinear} the solver and reconstruction were respectively downgraded to ``\texttt{HLL}'' and ``\texttt{PARABOLIC}'' for numerical stability purposes. The $\nabla \cdot \mathbf{B}=0$ condition was controlled using the ``\texttt{CONSTRAINED$\_$TRANSPORT}'' method under the ``\texttt{UCT$\_$HLL}'' scheme. ``\texttt{ENTROPY$\_$SWITCH}'' was always kept off.

\begin{figure*}[t]
    \centering
    \includegraphics[width=18cm]{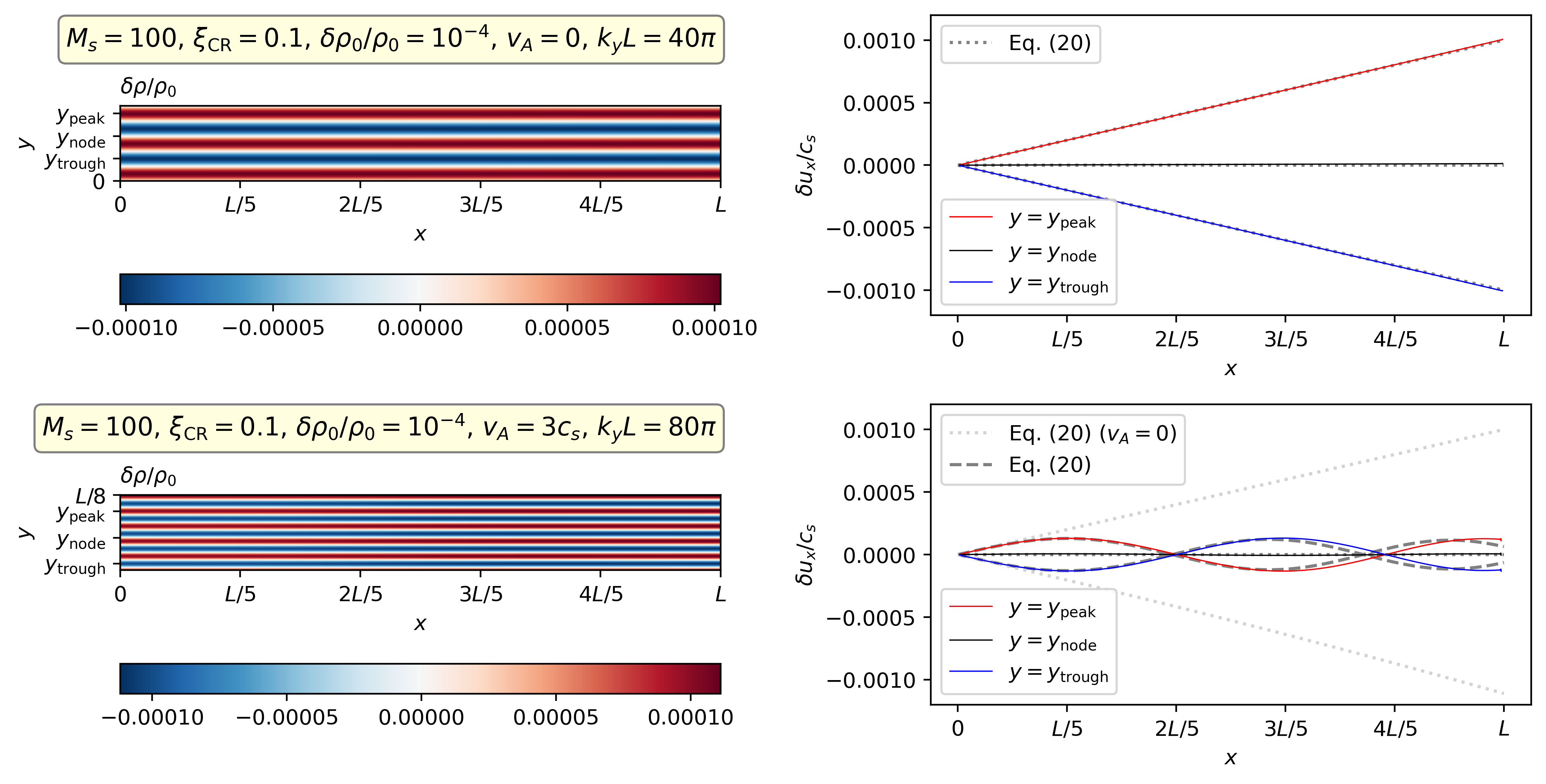}
    \caption{Stationary profiles of the dimensionless density (left) and velocity (right) perturbations from $k_y$ modes. On the right, red/blue/black lines show the simulation results for $\delta u_x/c_s$ at a fixed $y$ coordinate corresponding to the peak/trough/node of $\delta\rho/\rho_0$ oscillations. Dotted and dashed lines correspond to the analytical predictions of Eq.~\eqref{ky-withB}. \textit{Top}: No magnetic fields. \textit{Bottom}: Magnetic field perpendicular to the shock normal.}
    \label{fig:ky}
\end{figure*}

\subsection{The linear regime}\label{sec:linear}

We consider the linear regime where the perturbations produced by AI are much smaller than the unperturbed quantities in the entire simulation box. At $x=0$, the BC is set to:
\begin{equation}\label{eq:BC}
    \rho(x=0,y,t) = \rho_0+\delta\rho_0\sin(k_x u_0 t + k_y y + \phi)~,
\end{equation}
where $0\leq\phi <2\pi$ is a random phase, $\mathbf{u}=u_0\, \hat{\mathbf{x}}$, $P=P_0$, and $\mathbf{B}=B_{x,0}\,\hat{\mathbf{x}}$ or $\mathbf{B} = B_{y,0}\,\hat{\mathbf{y}}$. Eq.~\eqref{eq:BC} corresponds to a continuous inflow of density fluctuations with wavelengths $\lambda_x=2\pi/k_x$ and $\lambda_y=2\pi/k_y$ along the $x$ and $y$ directions. After one advection time, all transients due to initial conditions and boundary discontinuities at $t=0$ disappear and the system reaches a quasi-stationary state. In Sects.~\ref{sec:kx} and \ref{sec:ky} we study the behavior of $k_x$ modes (where $k_y=0$) and $k_y$ modes (where $k_x=0$) and test the growth rates of AI derived in Sect.~\ref{sec:instability}. In Sect.~\ref{sec:manymodes} we investigate the more realistic scenario with many $k_x$ and $k_y$ modes.

\subsubsection{Single $k_x$ mode}
\label{sec:kx}

The simulations in this section were performed on a default grid resolution of $N_x=4000$ and $N_y=4$. The chosen value of $N_x$ is large enough to avoid significant numerical damping. Since $\nabla P_{\rm CR}$ is oriented along the $x$-axis, density perturbations with $k_y=0$ are expected to induce $\delta u_x$, $\delta P$ and $\delta B_y$ perturbations that grow due to AI. In order to study the evolution of perturbations, we subtract the background profiles from the total quantities provided as output by PLUTO, for example
\begin{align}
    &\delta\rho(x,y,t) = \rho(x,y,t)-\frac{\rho_0}{U(x)}~,\label{eq:deltarho}\\
    &\delta u_x(x,y,t) = u_x(x,y,t) - u_0U(x)~.\label{eq:deltaux}
\end{align}

To assess the validity of the analytic results derived in Sect.~\ref{sec:instability}, we performed a scan over many different values of the parameters $\xi_{\rm CR}$, $M_s$, $\delta\rho_0/\rho_0$, $v_{\rm A}/c_s$, and $k_xL=2\pi \lambda_x^{-1}L$. In Fig.~\ref{fig:kx-steady-states}, we show some selected simulation results. Each subplot depicts constant-$y$ slices of density (black), pressure (red), velocity (blue), and magnetic field (green) perturbations at steady state, along with the corresponding expected growths $\propto\exp[\Gamma t(x)]$ (dashed lines), which are all consistent with the simulation results\footnote{An initial transient is present since we inject pure density perturbations, which are not eigenmodes of the instability. This transient is particularly visible in the top-right and bottom-left panels of Fig.~\ref{fig:kx-steady-states}. Once velocity/pressure/magnetic field fluctuations develop, the exponential growth becomes clear.}. In the top-left panel, we adopt $\xi_{\rm CR}=0.1$, $M_s=100$, $\delta\rho_0/\rho_0 = 10^{-4}$, $k_xL=80\pi$, and no background magnetic field, while in the bottom-left panel we add a magnetic field perpendicular to the shock normal. As expected, a perpendicular magnetic field suppresses the growth rate of the instability. We also confirmed that a parallel field $\mathbf{B}_0\parallel\hat{\mathbf{x}}$ reproduces the zero-field result. In the top-right panel, we probe the small-wavelength limit ($k\gtrsim |\nabla P_{\rm CR}|/\rho_0c_{\rm s}^2=\xi_{\rm CR}M_s^2/L$), which requires a factor-10 increase in grid resolution ($N_x=40000$) to avoid numerical dissipation. Finally, in the bottom-right panel, we consider larger values of $\delta\rho_0/\rho_0$ and $\xi_{\rm CR}$. In this case, the amplitude of the perturbations approaches the nonlinear regime close to the shock. As we shall discuss in Sect.~\ref{sec:nonlinear}, in this regime, the perturbations steepen and eventually lead to the formation of shocklets.

\begin{figure*}
    \centering
    \includegraphics[width=18cm]{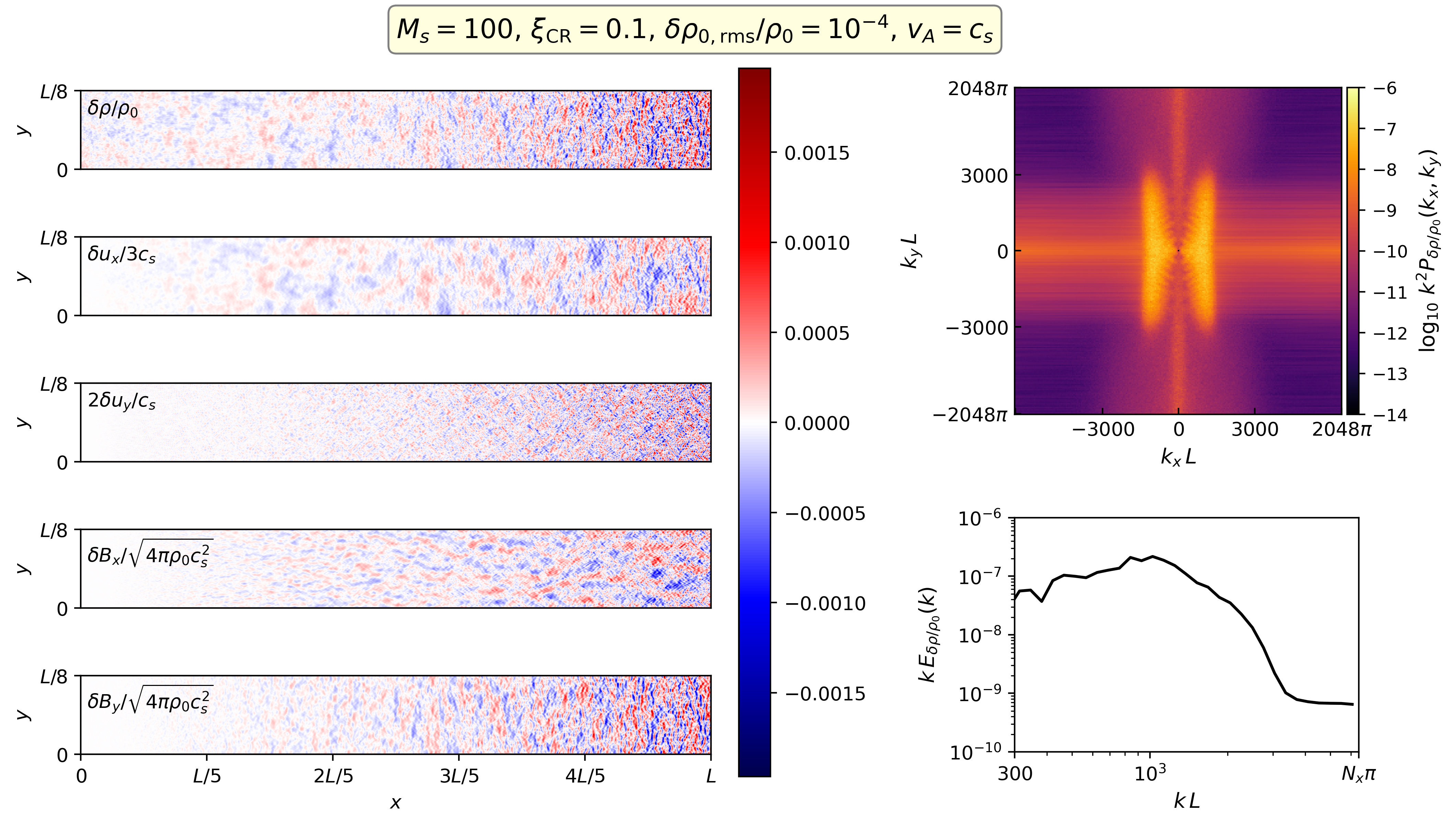}
    \caption{Linear-regime simulation results for initial density fluctuations following Eq.~(\ref{eq:DDdensity}) with a flat power spectrum in a $2048\times256$ grid. \textit{Left}: Perturbation profiles for (from top to bottom) density, velocity $x$- and $y$-components, and magnetic field $x$- and $y$-components. Note that $\delta u_x/c_s$ has been multiplied by $1/3$ and $\delta u_y/c_s$ by $2$, so all quantities can be visualized using the same color bar scale. \textit{Right}: dimensionless 2D (top) and omnidirectional (bottom) density fluctuations power spectra, $k^2 P_{\delta\rho/\rho_0}(k_x,k_y)$ and $kE_{\delta\rho/\rho_0}(k)$, respectively.}
    \label{fig:many-modes}
\end{figure*}

\subsubsection{Single $k_y$ mode} \label{sec:ky}

The dispersion relations given by Eqs.~\eqref{eq:kperp-dispersion} and \eqref{eq:kpar-dispersion} show that only perturbation wave vectors parallel to $\nabla P_{\rm CR}$ lead to instability. However, perturbations with $k_x=0$ are also affected by the CR pressure gradient because the acceleration $-\nabla P_{\rm CR}/\rho$ of a fluid element depends on the local density (overdensities decelerate slightly less than underdensities). For density perturbations along $y$, one has
\begin{equation}\label{eq:relative-acceleration}
    -\frac{\nabla P_{\rm CR}}{\rho_0 + \delta\rho_0\sin(k_y y)} \approx -\frac{\nabla P_{\rm CR}}{\rho_0} + \frac{\nabla P_{\rm CR}}{\rho_0}\frac{\delta\rho_0}{\rho_0}\sin(k_y y)~.
\end{equation}
The first term corresponds to an overall deceleration of the fluid, while the second term leads to a relative acceleration of overdensities with respect to underdensities. The evolution of $\delta u_x$ is given by (see Appendix \ref{app:ky-velocity})
\begin{equation}\label{ky-withB}
    \frac{\delta u_x(x,y)}{u_0} \approx \frac{\nabla P_{\rm CR}}{\rho_0u_0^2}\frac{\delta\rho_0}{\rho_0}\,\sin(k_y y)\, \frac{\sin(k'_x x)}{k'_x} ~,
\end{equation}
where
\begin{equation}
k'_x = k_y\sqrt{\frac{B_{y,0}^2}{4\pi\rho_0u_0^2 U^3}}=\frac{k_y}{M_{A,y} U^{3/2}} ~.
\end{equation}
Eq.~(\ref{ky-withB}) can be interpreted as follows. Underdensities are more decelerated by the cosmic ray pressure gradient with respect to overdensities. Magnetic field lines bend because they are frozen in the fluid, and magnetic tension acts as a restoring force. Then, $\delta u_x$ oscillates while the fluid moves towards the shock. When there is no magnetic field, $\delta u_x$ increases linearly.

This effect can be seen in Fig.~\ref{fig:ky}, which shows the results of simulations performed on a grid with $N_x=2000$ and $N_y=250$. The top panels show the steady-state result of a simulation with the same parameters as in the top-left panel of Fig.~\ref{fig:kx-steady-states}, but with a $k_y=40\pi$ mode instead of a $k_x=80\pi$ mode. On the left, we show $\delta\rho/\rho_0$. Sinusoidal oscillations along the $y$-axis do not grow exponentially, as expected. On the right, we show $\delta u_x$. We consider different $y$ slices: $y=y_{\rm peak}$ (red), which is a peak of $\delta\rho/\rho_0$; $y=y_{\rm trough}$ (blue), which is a trough; and $y=y_{\rm node}$ (black), which is a node. The evolution closely matches the prediction of Eq.~\eqref{ky-withB} also when $B_{y,0}=0$, shown as dotted lines.

The same quantities are shown in the bottom panels of Fig.~\ref{fig:ky}, but for a different choice of parameters -- notably, with the addition of a perpendicular magnetic field such that $v_{\rm A}=3c_{\rm s}$. The density profile looks qualitatively similar to the one in the top panel, indicating that no significant forces arise along the $y$ direction. However, the velocity perturbations oscillate around zero, rather than growing linearly, because of magnetic tension. We verified the appearance of $\delta B_x$ in this simulation, thereby confirming our interpretation.

\subsubsection{Many modes}
\label{sec:manymodes}

\begin{figure*}[t]
    \centering
    \includegraphics[width=18cm]{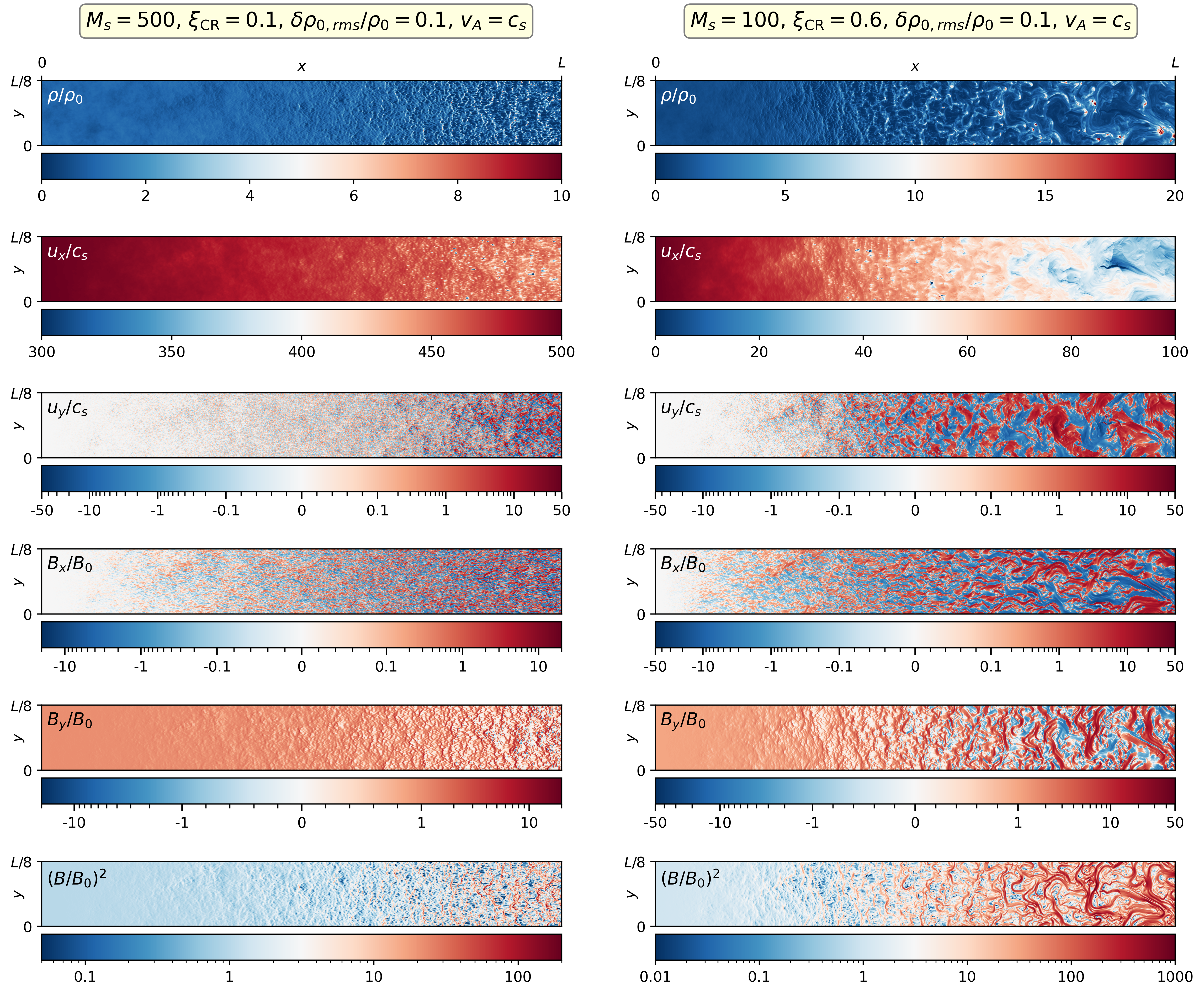}
    \caption{2D snapshots of quasi-stationary density, velocity, magnetic field, and magnetic energy density profiles, for different parameter choices, as indicated above each column, in a $4096\times512$ simulation grid. All quantities have been normalized so as to become dimensionless. For better contrast, colorbars use linear scaling for $\rho/\rho_0$ and $u_x/c_s$, logarithmic scaling for $(B/B_0)^2$, and symmetric logarithmic  scaling (linear near zero and logarithmic at large amplitudes) for other quantities.}
    \label{fig:nonlinear-profiles}
\end{figure*}

We now address the more realistic case of ISM inhomogeneities across multiple scales and along all directions. The injected density profile is
\begin{align}
    &\rho(x=0,y,t)=\rho_0\left[1 + \frac{\delta\rho_0}{\rho_0}(u_0 t,y)\right]~, \\
    &\frac{\delta\rho_0}{\rho_0}(x,y) = \sum_{\substack{a,b=0}}^{N_x,N_y} A_{ab}\cos\left[\frac{2\pi}{L} (ax+by) + \phi_{ab}\right]~,\label{eq:DDdensity}
\end{align}
where $A_{ab}$ and $\phi_{ab}$ are random amplitudes and phases of each perturbation. The latter are drawn uniformly from $0\leq \phi_{ab}<2\pi$, while the former are chosen in such a way to produce a flat omnidirectional power spectrum of ISM density fluctuations, $E_{\delta\rho_0/\rho_0}(k)\propto k^{-1}$, with a root-mean-square (RMS) amplitude $\delta\rho_{0,\rm rms}/\rho_0=10^{-4}$. This procedure is described in Appendix \ref{app:density}. Perturbations with wavelength $>L/8$ are set to have zero amplitude, while those with wavelength down to grid resolution scales are still injected into the system.

All power spectra in this work are calculated in the last eighth of the box (\textit{i.e.} $7L/8 <x < L$), since the amplitude of fluctuations is largest and nonlinear structures are most prominent in the part of the precursor that is closest to the shock. We also time-average the power spectra over several snapshots spanning one advection time, making them stationary.

Fig.~\ref{fig:many-modes} shows the steady-state $\delta\rho/\rho_0$, $\delta u_{x,y}$, and $\delta B_{x,y}$ profiles (left) and the corresponding $\delta\rho/\rho_0$ power spectra (right), from a simulation performed in a $2048\times 256$ grid using $\xi_{\rm CR}=0.1$, $M_s=100$, and $\mathbf{B}_0$ directed along $\hat{\mathbf{y}}$ such that $v_A=c_s$. The perturbations remain small before reaching the shock while still growing exponentially at the nominal AI rate. Tracking the growth of each individual mode along the box becomes impractical, as it requires the use of sophisticated techniques such as wavelet transforms \citep{2015JPlPh..81f4302F}. Here, we adopt the simple Fourier transform (FT), whose formalism is outlined in Appendix \ref{app:powerspec}, sacrificing spatial resolution in order to get the power of each $\mathbf{k}$ mode averaged over the whole box.

The top-right panel of Fig.~\ref{fig:many-modes} shows the 2D density power spectrum $P_{\delta\rho/\rho_0}(k_x,k_y)$, multiplied by $k^2$ to make it dimensionless. For $k\lesssim 10^3$, the observed power spectrum resembles the expected AI growth rate. In other words, the amplitude of each $(a,b)$ mode in Eq.~\eqref{eq:DDdensity} grows along the box as $A_{ab}(t)=A_{ab}\exp[\Gamma(k_a,k_b)\,t]$, where $\Gamma(k_a,k_b)$ is the AI growth rate obtained from Eq.~\eqref{eq:det0}. Numerical damping significantly affects the simulation results for $k\gtrsim 10^3$, evidenced by a clear break at $k\sim 10^3$ in the time-averaged omnidirectional power spectrum $E_{\delta\rho/\rho_0}(k)$ shown in the bottom-right panel. This is an unavoidable and critical feature of the simulations, as was also pointed out by DD and VLS.

\subsection{Nonlinear regime}
\label{sec:nonlinear}

\begin{figure*}[t]
    \sidecaption
    \parbox[t]{12cm}{
        \centering
        \includegraphics[width=\linewidth]{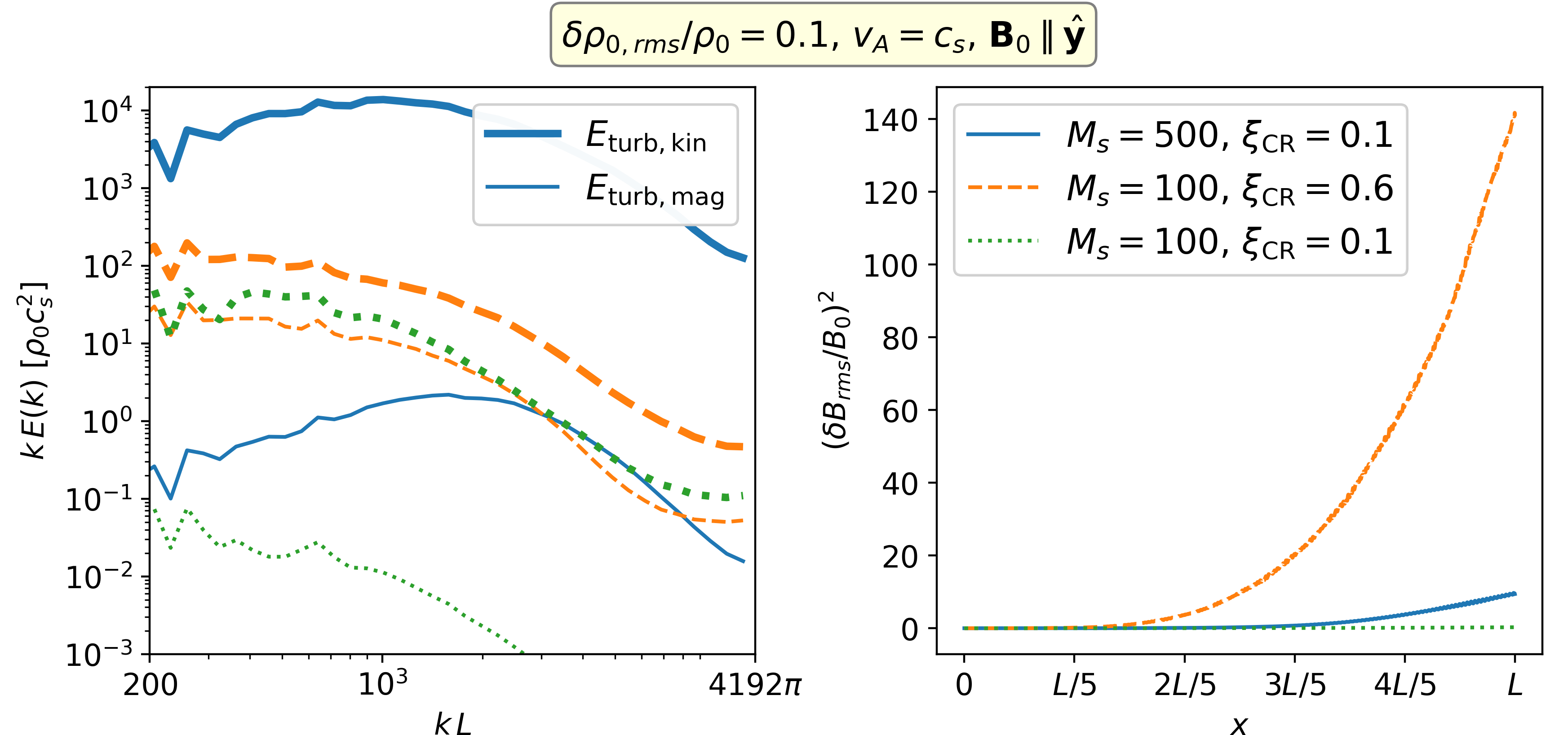}\\[0.3cm]
        \includegraphics[width=\linewidth]{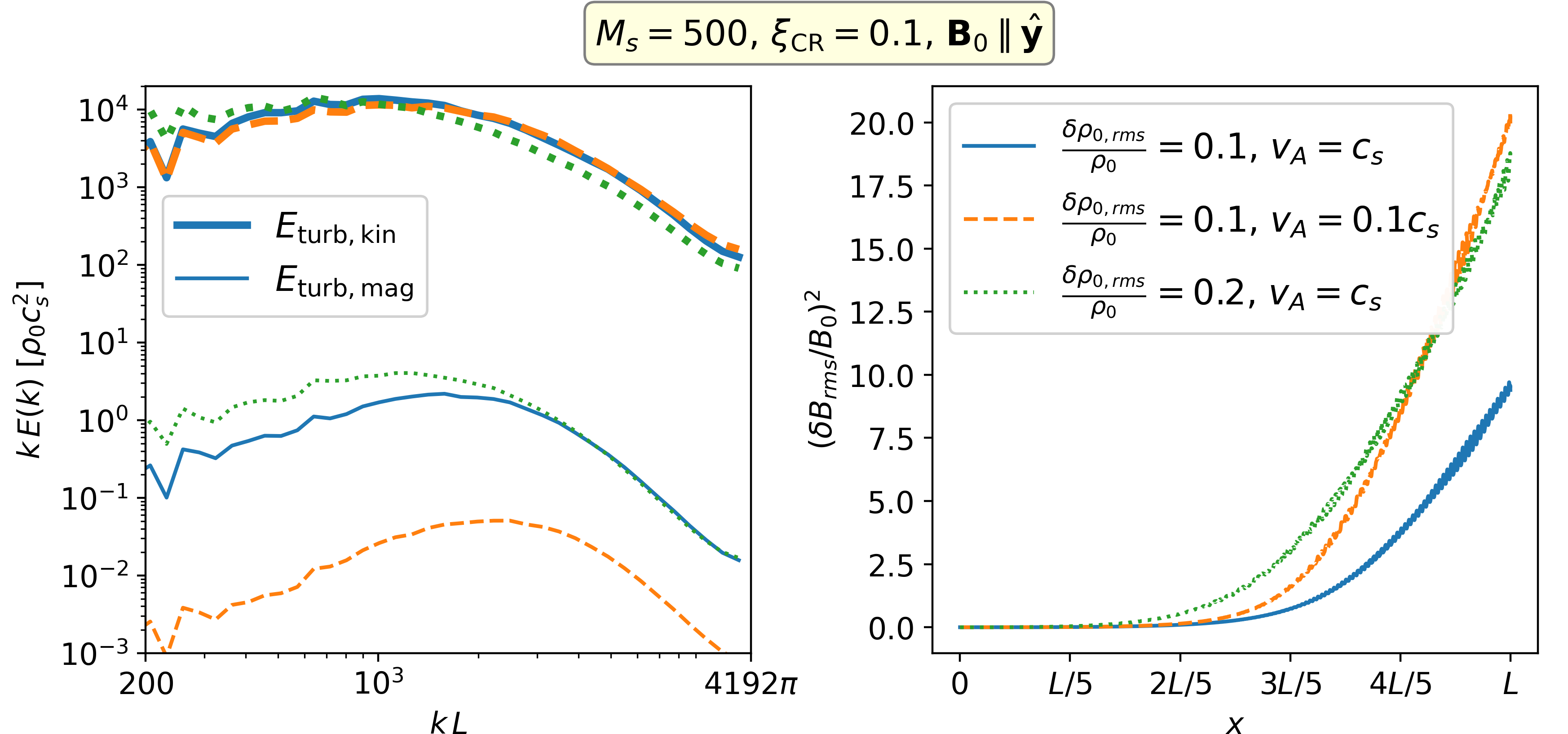}\\[0.3cm]
        \includegraphics[width=\linewidth]{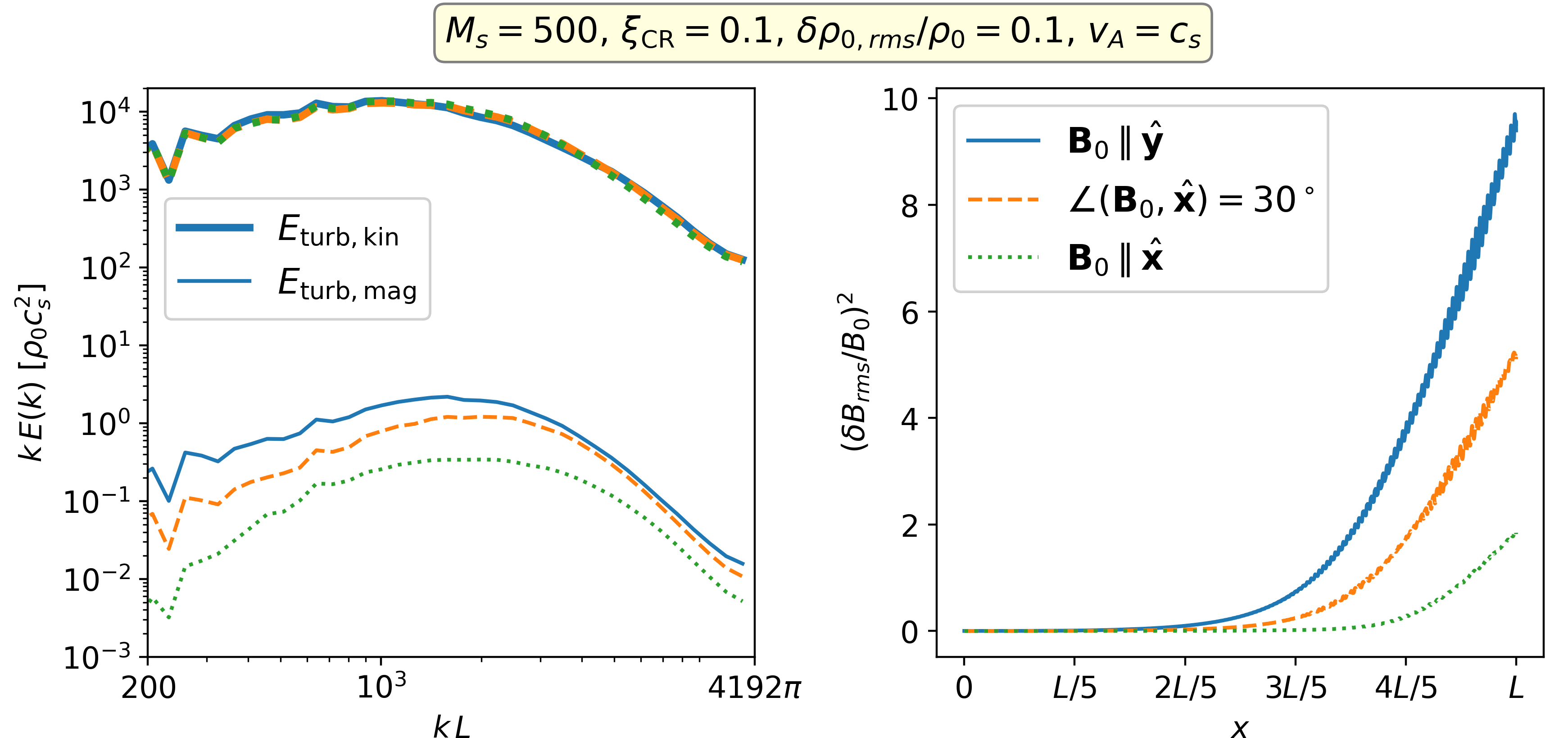}%
    }
    \caption{Turbulent power spectra in units of $\rho_0c_s^2$ (left) and MFA (right), calculated via Eq.~(\ref{eq:deltaBrms-growth}), for different parameter values and a $4096\times 512$ grid. Thick/thin lines in the left panels are kinetic/magnetic power spectra. The right panels' legends indicate the colors/line styles corresponding to each choice of parameters. \\
    \textit{Top}: we fix $\delta\rho_{0,rms}/\rho_0=0.1$, $v_A=c_s$, and $\mathbf{B}_0\parallel \hat{\mathbf{y}}$ and vary $M_s$ and $\xi_{\rm CR}$. \\
    \textit{Middle}: we fix $M_s=500$, $\xi_{\rm CR}=0.1$, and $\mathbf{B}_0\parallel \hat{\mathbf{y}}$ and vary $\delta\rho_{0,rms}/\rho_0$ and $v_A/c_s$. \\
    \textit{Bottom}: we fix $M_s=500$, $\xi_{\rm CR}=0.1$, $\delta\rho_{0,rms}/\rho_0=0.1$, and $v_A=c_s$ and vary the background field orientation.}
    \label{fig:nonlinear-parameters}
\end{figure*}

We consider the nonlinear regime where large perturbations are present near the shock surface ($\delta\rho,\delta P,\delta u, \delta B \gtrsim \rho_0,P_0,u_0,B_0$). Large $\delta B$ perturbations are expected to efficiently scatter particles, thus increasing their maximum energy. We show that AI can lead to large amplitude fluctuations even when the perturbations are small at upstream infinity. 

We employ the simulation framework described in Sect.~\ref{sec:manymodes}, injecting a superposition of perturbations down to the grid resolution scale. While the initial growth of pre-existing small perturbations at upstream infinity is due to the onset of AI in the CR precursor pressure gradient, their nonlinear evolution is connected with the onset of turbulence. A similar investigation was carried out by DD using 2D MHD simulations and by VLS using 3D simulations. Both DD and VLS adopted parameters aimed to maximize MFA, but not viable on physical grounds: for instance the efficiency of particle acceleration was $\xi_{\rm CR}=0.6$, which is now considered disproportionately large with respect to the results of hybrid simulations, as well as phenomenological considerations, suggesting $\xi_{\rm CR}\simeq 0.1$. In addition, previous studies considered Mach numbers $M_s\sim 100$, while shocks at the beginning of the Sedov phase of a SNR are expected to have $M_s\sim 500$, and even larger at earlier times.

The adoption of more realistic values of these parameters has profound consequences for MFA and for the onset of turbulence. In Fig.~\ref{fig:nonlinear-profiles}, we show 2D profiles of $\rho/\rho_0$, $u_{x,y}/c_s$, $B_{x,y}/B_0$ and $(B/B_0)^2$ for our benchmark choice of parameters ($\xi_{\rm CR}=0.1$, $M_s=500$, $\delta\rho_{0,rms}/\rho_0 = 0.1$, and $v_A=c_s$; left column), as well as for the parameters adopted in previous literature ($\xi_{\rm CR}=0.6$, $M_s=100$, $\delta\rho_{0,rms}/\rho_0 = 0.1$, and $v_A=c_s$; right column). The main qualitative differences are: 1) The left profiles contain fluctuating structures predominantly on smaller scales with respect to the right profiles. 2) The amplitude of $(\delta B/B_0)^2$ fluctuations is smaller on the left, indicating a lower level of MFA. 3) The combination of a longer advection time and a larger $\xi_{\rm CR}$ allows for early eddy formation in the right panels. Meanwhile, in the left panels, eddies form only very close to the shock, and most of the precursor is dominated by small-scale shocklets. \citet{Kang:1992vw} suggested that these shocklets can play a role in energizing particles, even before they cross the shock surface. In both cases, the morphology of $\delta B_x$ is elongated along $x$, while $\delta B_y$ is elongated along $y$ since the latter grows mainly due to plasma compression at the location of the shocklets.

\begin{figure*}
    \sidecaption
    \includegraphics[width=12cm]{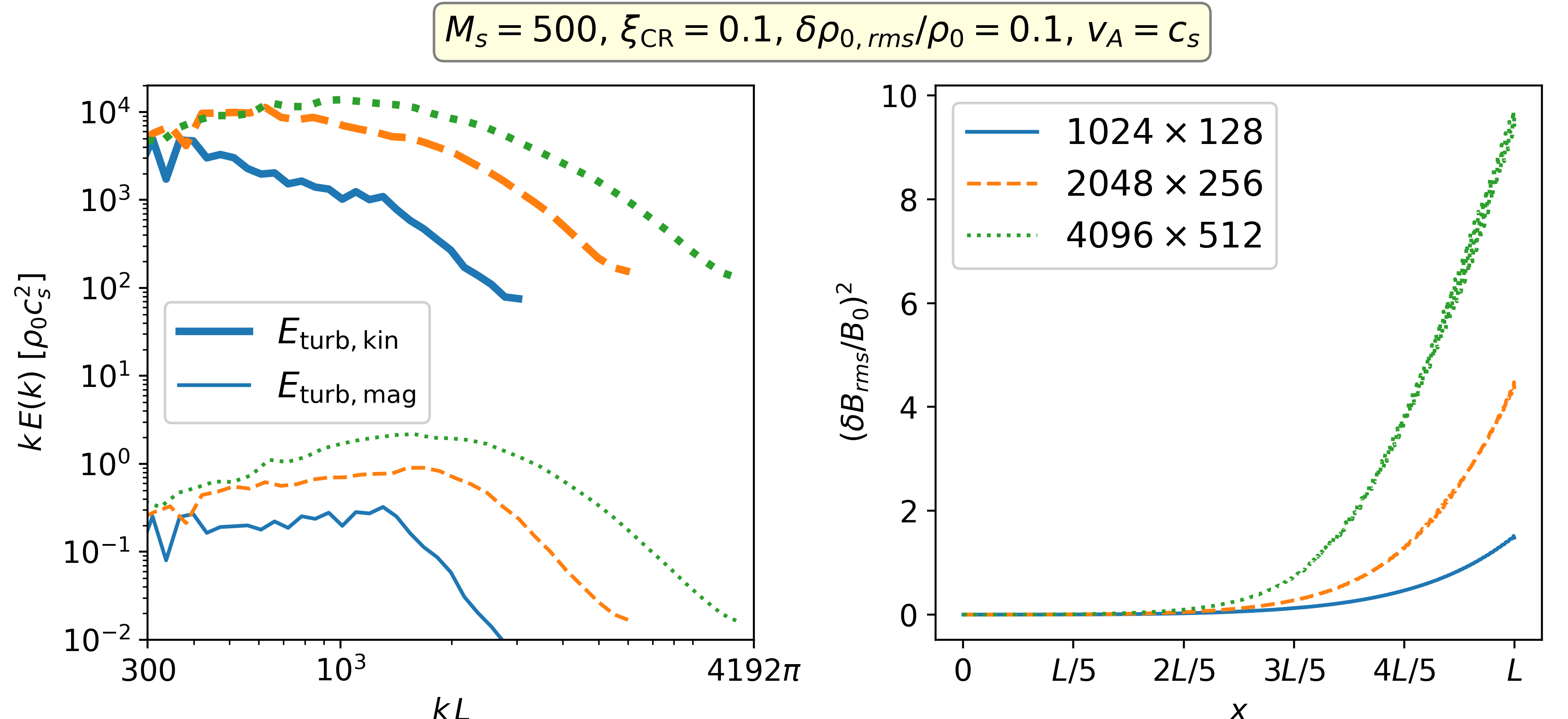}
    \caption{\textit{Left}: Turbulent kinetic (thick lines) and magnetic (thin lines) power spectra in units of $\rho_0c_s^2$ for different grid resolutions (see right panel's legend). \textit{Right}: MFA as a function of $x$ for different grid resolutions.}
    \label{fig:nonlinear-resolutions}
\end{figure*}

To quantitatively analyze our results, we calculate both kinetic and magnetic omnidirectional turbulent energy density spectra, $E_{\rm turb,kin}(k)$ and $E_{\rm turb,mag}(k)$ respectively, as defined in Appendix \ref{app:powerspec}. We first obtain the density-weighted velocity $\mathbf{w}(x,y)=\sqrt{\rho(x,y)}\,\mathbf{u}(x,y)$ (as is customary in studies of compressible MHD turbulence, \textit{e.g.} \citet{1990JSCom...5...85K,Grete:2017usl}) and the magnetic field $\mathbf{B}(x,y)$ from the simulation output, and then subtract their mean background profiles\footnote{We calculate $\overline{\mathbf{w}}$ and $\overline{\mathbf{B}}$ by averaging each field component over the $y$ (and $z$) direction(s) and over multiple snapshots in time, along an interval of $t_{\rm adv}$. This procedure ensures that $\delta\mathbf{w}$ and $\delta\mathbf{B}$ average to zero at each $x$.} $\overline{\mathbf{w}}(x)$ and $\overline{\mathbf{B}}(x)$ to obtain the turbulent components:
\begin{align}
    \delta \mathbf{w}(x,y) &= \mathbf{w}(x,y)-\overline{\mathbf{w}}(x)~,\\
    \delta \mathbf{B}(x,y) &= \mathbf{B}(x,y)-\overline{\mathbf{B}}(x)~.
\end{align}
The turbulent energy spectra are defined as the RMS energy density in turbulent fluctuations on scales between $k$ and $k+dk$:
\begin{align}
    E_{\rm turb,kin}(k)\,dk &= \frac{\delta w_{rms}^2(k)}{2}~,\\
    E_{\rm turb,mag}(k)\,dk &= \frac{\delta B_{rms}^2(k)}{8\pi}~.
\end{align}
We also calculate the strength of MFA as a function of the distance from the shock surface:
\begin{equation}\label{eq:deltaBrms-growth}
    \frac{\delta B^2_{rms}(x)}{B^2_0}~, \quad \delta B_{rms}^2(x) = \frac{1}{L/8}\int_0^{L/8}|\delta \mathbf{B}(x,y)|^2\, dy~.
\end{equation}

In Fig.~\ref{fig:nonlinear-parameters}, we show the turbulent power spectra and MFA for different choices of parameters ($M_s$, $\xi_{\rm CR}$, $v_A/c_s$, $\delta\rho_{0,rms}/\rho_0$, and magnetic field orientation). In the top panels, we fix $\delta\rho_{0,rms}/\rho_0=0.1$, $v_{\rm A}=c_s$, and $\mathbf{B}_0\parallel\hat{\mathbf{y}}$ while varying the remaining parameters. We compare the results for the benchmark values of the parameters (blue solid) with the DD/VLS parameters (yellow dashed) and with the more conservative case $M_s=100$, $\xi_{\rm CR}=0.1$ (green dotted). It is worth recalling that for a typical SNR shock, the Mach number in the phase that is expected to be most important for particle acceleration is typically $\sim 100-1000$ and the efficiency is $\xi_{\rm CR}\sim 0.1$. From the top-right panel we immediately see that the scenario with $M_s=100$ and $\xi_{\rm CR}\sim 0.1$ does not yield any appreciable MFA; the background field still carries more energy than the fluctuations in this case. Note that in our benchmark case the turbulent magnetic energy spectrum is very hard, with most of the power concentrated at small scales, which could translate into enhanced scattering of particles at the energies of interest (see discussion in Sect.~\ref{sec:powerspec}). Meanwhile, strongly modified shocks with $\xi_{\rm CR}=0.6$ can easily produce $\delta B_{rms}/B_0 \gtrsim 10$, but their turbulent magnetic energy spectra are somewhat steep, with most power concentrated on large scales. This is the only scenario in which we are anywhere close to reaching energy equipartition:
\begin{equation}\label{eq:equipartition}
    E_{\rm turb,\,kin}(k)\approx E_{\rm turb,\,mag}(k)~,\quad {\rm for}~k>k_{\rm eq}~,
\end{equation}
where $k_{\rm eq}^{-1}$ is the equipartition length scale. For smaller $\xi_{\rm CR}$, the Mach number must be large to obtain any meaningful MFA. But increasing the Mach number also increases $E_{\rm turb,\,kin}$ and $k_{\rm eq}$. Ultimately, it is hard to estimate $k_{\rm eq}$ from simulations because this scale is never numerically resolved and $E_{\rm turb,\,mag}(k)$ does not follow the $\sim k^{-1/2}$ scaling predicted by \citet{Beresnyak:2009pi}.

In the middle panels of Fig.~\ref{fig:nonlinear-parameters} we instead fix $M_s=500$ and $\xi_{\rm CR}=0.1$ and vary the remaining parameters while keeping $\mathbf{B}_0\parallel\hat{\mathbf{y}}$. For the yellow dashed curves, we set $v_A=0.1c_s$, while for the green dotted curves, we double the RMS amplitude of the initial fluctuations, $\delta\rho_{0,rms}/\rho_0=0.2$. Inspecting the low-$k$ region of the power spectra, where numerical damping is negligible, we find that increasing $v_A$ by a factor 10 increases $E_{\rm turb,\,mag}$ by a factor $\approx 100$. Similarly, doubling $\delta\rho_{0,rms}/\rho_0$ makes $E_{\rm turb,\,mag}$ larger by a factor $\approx 4$. This is consistent with the natural expectation that $E_{\rm turb,\,mag}\propto B_0^2(\delta\rho_{0,rms}/\rho_0)^2$.

\begin{figure*}
    \centering
    \includegraphics[width=\linewidth]{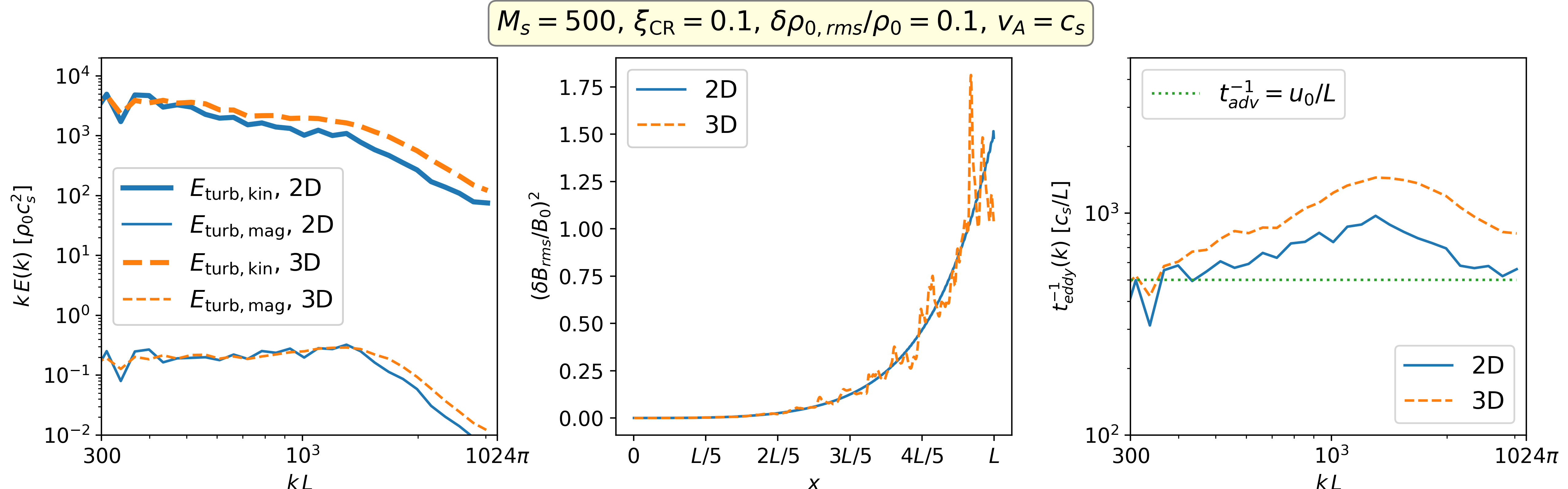}
    \caption{\textit{Left}: Turbulent kinetic (thick lines) and magnetic (thin lines) power spectra in units of $\rho_0c_s^2$ for 2D ($1024\times 128$ grid; blue solid) and 3D ($1024\times 128\times 128$ grid; yellow dashed) simulations. \textit{Center}: MFA as a function of $x$ for 2D and 3D simulations. In 3D, we perform a $yz$-averaging rather than just a $y$-averaging. \textit{Right}: Inverse eddy turnover timescales for the 2D (blue solid) and 3D (yellow dashed) runs, compared to the inverse advection timescale $t_{\rm adv}^{-1}=u_0/L$ (dotted green), in units of $c_s/L$.}
    \label{fig:nonlinear-dimensions}
\end{figure*}

In the bottom panels we keep $M_s$, $\xi$, $\delta\rho_{0,rms}/\rho_0$, and $v_A/c_s$ fixed to their benchmark values to isolate the effect of field orientation. The results are somewhat consistent with the interpretation put forward by \citet{Downes:2014rta}, who postulated that amplification occurs primarily to the component $\delta B_y$. If the shock is parallel ($\mathbf{B}_0\parallel \hat{\mathbf{x}}$, green dotted lines), magnetic fluctuations along $y$ arise only after the system turns nonlinear. Instead, if $\mathbf{B}_0\parallel \hat{\mathbf{y}}$ (blue solid lines), $\delta B_y$ is immediately generated in AI eigenmodes. The latter is the most optimistic scenario for MFA. However, more generally, SNR shocks expand onto oblique fields and MFA saturates between the two extreme cases discussed above. As an example, we show the case in which $\mathbf{B}_0$ is inclined by $30^\circ$ with respect to the shock normal (yellow dashed lines). At low-$k$, its magnetic power spectrum is $\sim 4$ times lower than the perpendicular shock case. This indicates that  $\delta B_y\propto \mathbf{B}_0\cdot\hat{\mathbf{y}}=B_0\sin30^\circ=B_0/2$ indeed dominates the amplified field.

To meaningfully interpret these results,  in Fig.~\ref{fig:nonlinear-resolutions} we perform a numerical convergence test, adopting the benchmark choice of parameters. In the left panel, we show the turbulent kinetic (thick lines) and magnetic (thin lines) power spectra, for different grid resolutions, normalized to $\rho_0 c_s^2$. The magnetic power peaks at larger $k$ for higher resolution, but is always far from reaching equipartition with the kinetic power. Note that doubling the grid resolution doubles the wavenumber $k_d$ above which damping is important ($k_dL\approx 1.3\times 10^3$ for the $1024\times 128$ resolution) and changes the shape of $E_{\rm turb,\,mag}(k)$ below $k_d$, making it harder. This is not only due to the reduced numerical damping for $k\lesssim k_d$, but also happens because less power can be generated and transferred from higher to smaller $k$ via inverse cascading. In the right panel, we show $(\delta B_{rms}/B_0)^2$ as a function of $x$. As we can see, MFA increases with resolution, without converging. Our $\delta B/B_0\sim$~a few can only be interpreted as a conservative lower limit, because most of the magnetic energy is expected to be located at damping-dominated scales $kL \gtrsim 2000$. As we will see in Sect. \ref{sec:powerspec}, these scales play a crucial role for particle scattering.

In this work, we focused on 2D simulations in order to access smaller scales, which are the most important because they carry most of the turbulent magnetic energy. Both DD and VLS have shown that simulations in 2D yield less power on small scales than in 3D. This is likely because of turbulent kinetic energy predominantly undergoes an inverse cascade in 2D and a forward cascade in 3D. However, for this difference to be appreciable, turbulence should develop and have enough time to transfer kinetic energy between scales. Energy can safely cascade between $k_1\to k_2$ when the eddy turnover time $t_{\rm eddy}(k) \sim 1/k\,\delta u(k)$ at all intermediate wavenumbers $k_1<k<k_2$ is much shorter than the advection time in the precursor:
\begin{equation}\label{eq:eddy-turnover}
    k\,\delta u_{rms}(k) \gg t_{\rm adv}^{-1}=u_0/L~, \quad {\rm for} \quad k_1<k<k_2~.
\end{equation}
The RMS amplitude of velocity fluctuations at a given $k$, $\delta u_{rms}(k)$, is defined via $E_{\delta u}(k)\,dk=\rho_0\delta u^2_{rms}(k)/2$, where the velocity power spectrum $E_{\delta u}$ is defined in Appendix \ref{app:powerspec}. Note that $E_{\delta u}$ is different from $E_{\rm turb,\,kin}$, which is defined in terms of $\mathbf{w}=\sqrt{\rho}\mathbf{u}$.

In Fig.~\ref{fig:nonlinear-dimensions}, we compare power spectra (left panel), MFA (middle panel), and characteristic timescales (right panel), for our benchmark set of parameters and $1024\times 128~(\times 128)$ resolution, in 2D (blue solid lines) and 3D (yellow dashed lines) simulations respectively. Indeed, we find slightly more kinetic power at higher $k$ in the 3D scenario, as well as comparable MFA. However, numerical damping once again prevents us from resolving the scales in which this difference could become appreciable. By comparing the eddy turnover times in the right panel with the advection time (green dotted line), we see that condition (\ref{eq:eddy-turnover}) is only marginally satisfied, and significant energy cascading is not expected to take place. In contrast, using the overly optimistic $\xi_{\rm CR}=0.6$ and $M_s=100$ adopted in DD and VLS, one expects stronger cascading simply because $\Gamma\,t_{\rm adv}$ is larger in the $kL\ll \xi_{\rm CR}M_{ms}^2$ limit, thus reaching the nonlinear/turbulent regime earlier. In summary, while 2D allows us to access perturbations with large wavenumber, 3D simulations may provide a better description of the nonlinear phase if small scales are resolved. On the other hand, it is clear that accessing such scales in 3D simulations becomes quickly unbearably expensive from the computational point of view.

\section{Competition and cooperation between acoustic and non-resonant instability}
\label{sec:powerspec}

Particle acceleration at SNR shocks must be accompanied by substantial MFA, not only to reach a higher value of the maximum energy, but also to explain the thin filaments observed in X-rays \cite[]{Vink2012}. It is however important to keep in mind that while the former point requires MFA upstream of the shock, namely in the CR precursor, to reduce the acceleration time, the latter only requires magnetic field enhancement behind the shock. 

All processes that can in principle lead to MFA are associated with a gradient in the CR distribution upstream: in the case of the non-resonant instability \cite[]{Bell:2004hhd}, it is the current of particles escaping the upstream region that leads to the growth of the instability. In general, this phenomenon occurs on scales larger than the precursor size, which is $L\sim D(E_{\rm max})/v_{\rm sh}$, where $D$ is the diffusion coefficient and $E_{\rm max}$ is the maximum energy of the accelerated particles (for convenience, we use $v_{\rm sh}$ rather than $u_0$ in this section). Such current is directly related to the gradient of CR number density and the process may lead to reaching $\delta B/B\gg 1$. For the resonant streaming instability \cite[]{Lagage:1983zz}, it is the gradient of the CR density upstream to produce Alfv\'en waves at wavenumbers resonant with the CR momentum ($r_L(E)\simeq k^{-1}$). This latter process in general saturates at $\delta B/B\lesssim 1$, especially if some type of damping is effective. Finally, AI discussed in this article grows due to the dynamical action of the CR pressure gradient upstream of the shock on density fluctuations present in the ISM and can lead to $\delta B/B\gg 1$, depending on whether the instability has enough time to grow in one advection time. Below we comment on the location where these instabilities can be excited, on the scales involved and which one is expected to dominate as a function of the parameters.  

If, for simplicity, we assume that the spectrum of accelerated particles in a power law in momentum space $f(p)\propto p^{-4}$, appropriate for strong shocks, then the condition for the growth of the non resonant instability can be written as:
\begin{equation}
    \frac{3 \xi_{\rm CR}\rho_0 v_{\rm sh}^2}{\Lambda}\frac{v_{\rm sh}}{c} \geq \frac{B_0^2}{4\pi},
    \label{eq:threshold}
\end{equation}
where $\Lambda\approx \ln( E_{\rm max}/m_p c^2)\sim 10$. This condition is equivalent to requiring that the energy density carried by the current is larger than the energy density in the pre-existing magnetic field \cite[]{Bell:2004hhd,AmatoBlasi2009}. If one introduces the Alfvénic Mach number in the pre-existing field, then this condition can be rewritten as
\begin{equation}
    M_A\geq \left( \frac{c \Lambda}{3 \xi_{\rm CR} v_A}\right)^{1/3}\approx 100\, \left( \frac{\xi_{\rm CR}}{0.1} \right)^{-1/3} \left( \frac{v_A}{10 \rm\; km/s} \right)^{-1/3}.
\end{equation}
Typically, when the shock velocity drops below $\sim 1000$ km/s, the non-resonant instability stops being excited and one has only the resonant streaming instability to rely upon. 

At saturation, the magnetic field $B_{\rm sat}$ can be estimated from Eq.~\eqref{eq:threshold} by replacing $B_0$ with $B_{\rm sat}$ and interpreting it as an equality. Following \citet[]{Schure2013,Cristofari2020}, the maximum energy is obtained by requiring that the instability grows for $n$ e-folds, where typically it is assumed that $n\approx 5$, namely $\gamma_{\rm max} t\approx 5$, where $\gamma_{\rm max}$ is the maximum growth rate of the instability. When the non resonant instability is excited, it is therefore driven by the current of escaping particles over a region with size $L_{MFA}\sim 5c \gamma_{\rm max}^{-1}$, where
$\gamma_{\rm max}=k_{\rm max} v_A$ and 
\begin{equation}
    k_{\rm max}=\frac{4\pi}{c B_0}J_{\rm CR}(>E)=\frac{4\pi}{c B_0}   \frac{3 \xi_{\rm CR} \rho_0 v_{\rm sh}^2}{\Lambda E}e v_{\rm sh}
\end{equation}
is the scale where the growth is the fastest \cite[]{Bell:2004hhd}. Here $J_{\rm CR}(>E)$ is the current carried by particles with energy $>E$. It can be easily shown that the condition in Eq.~\eqref{eq:threshold} is also equivalent to requiring $k_{\rm max}\geq r_L(E)^{-1}$, where $r_L(E)$ is the Larmor radius of the particles with energy $E$. The region affected by MFA, $L_{MFA}$, is typically much larger than the size of the precursor, and the amplification of the magnetic field occurs on such a larger region. It is also important, for the considerations below, to keep in mind that the instability also results in overdensities $\delta \rho/\rho_0\gtrsim 1$ in the same region where the field is amplified. 

At saturation, the non resonant instability produces magnetic structures on the scale of the Larmor radius of the escaping particles in the amplified magnetic field. Using the considerations above, one finds that, at the highest energy $E_{\rm max}$, the Larmor radius in the amplified field is
\begin{equation}
    r_L(E_{\rm max}) = \frac{1}{5}\left(\frac{3\xi_{\rm CR}}{\Lambda}\right)^{1/2} \left( \frac{v_{\rm sh}}{c}\right)^{1/2} v_{\rm sh} t.
\end{equation}
At this point, the size of the precursor would be of the order $L\simeq D(E_{\rm max})/v_{\rm sh}$, corresponding to a scale $L r_{\rm L}^{-1}\simeq c/3 v_{\rm sh}$ in the Bohm limit. For $v_{\rm sh}=10^3-10^4$ km/s this corresponds to $L r_{\rm L}^{-1}\simeq 100-10$. Notice that these latter considerations apply to the non-resonant instability at saturation, while the wavenumber where initially (in the linear regime) the instability grows the fastest, $k_{\rm max}$, corresponds to much smaller scales. 

It is also useful to estimate $L/R_{\rm sh}$, where $R_{\rm sh}$ is the radius of the shock, which is obtained by approximating $v_{\rm sh}t\approx R_{\rm sh}$:
\begin{equation}
    \frac{L}{R_{\rm sh}}\approx \frac{1}{15}\left(\frac{3\xi_{\rm CR}}{\Lambda}\right)^{1/2} \left( \frac{v_{\rm sh}}{c}\right)^{-1/2}. 
\end{equation}
For $v_{\rm sh}=10^3-10^4$ km/s this corresponds to $L R_{\rm sh}^{-1}\simeq 0.2-0.06$, consistent with the typical approximation that $D(E_{\rm max})/v_{\rm sh}\sim 0.1 R_{\rm sh}$. 

The simple estimates presented here are formally valid for a parallel shock, although \cite{Bell2005} showed that similar results are obtained for oblique shocks. Hence we will adopt the estimate of $L$ provided above for oblique shocks as well. For shocks where the inclination to the shock normal is $\gtrsim 45^\circ$, injection has been shown to be suppressed \cite[]{Caprioli2014} and they are of lesser interest here.

Let us now estimate these same quantities for AI. The fastest growing modes grow at a rate $\sim \xi_{\rm CR}M_{ms} v_{\rm sh}/2 L$, which is larger than the non resonant growth rate at $E_{\rm max}$ when $M_{ms}\gtrsim (10/\xi_{\rm CR})(L/R_{\rm sh})\sim 10$, where the last inequality holds when we adopt $\xi_{\rm CR}$ and $L/R_{\rm sh}\sim 0.1$. Based on these considerations, AI seems to grow faster for the high Mach number shocks that are typical of young SNRs. 

As discussed above, the saturation of the instability is not easily accessible to current simulations, in that it requires the small scale magnetic modes to reach some sort of equipartition with kinetic perturbations and initiate a nonlinear cascade toward larger scales. If to trust the qualitative expectations of DD, one could suggest that at saturation
\begin{equation}
    \frac{\delta B^2}{8 \pi} \approx \xi_{\rm CR}^2 \rho_0 v_{\rm sh}^2 \left( \frac{\delta \rho}{\rho_0}\right)^2,
    \label{eq:DD}
\end{equation}
and assuming $\delta\rho/\rho_0\sim 1$, which is the result of the nonlinear growth of the instability, we get the following ratio between the saturation values of the acoustic/turbulent to non-resonant instability:
\begin{equation}
\frac{(\delta B^2)_{\rm ai}}{(\delta B^2)_{\rm nr}}\approx \frac{2\xi_{\rm CR}\Lambda}{3}\frac{c}{v_{\rm sh}}.    
\end{equation}
For typical values of the parameters, this ratio exceeds unity, which suggests that AI may prevail. As stressed above, numerical simulations of AI are not yet suitable to really test this regime and check the viability of Eq.~\eqref{eq:DD}.

We want to stress that the two CR induced instabilities studied here may operate at the same time and in fact cooperate rather than competing with each other: the fact that the Bell instability operates due to the current of particles escaping the acceleration region implies that this instability produces density perturbations $\delta \rho/\rho_0\gtrsim 1$ upstream of a shock, over a region much larger than the size of the precursor. When these regions enter the shock precursor, they can undergo additional MFA due to the onset of AI and its nonlinear counterpart. 

However, this conclusion requires some care: AI is expected to amplify the perpendicular component of the magnetic field, while, as discussed above, MFA is reduced in the case of parallel shocks. On the other hand, it is well known that the non-resonant instability creates large perpendicular magnetic fields away from the shock. Hence by the time that these perturbations enter the precursor they can further be amplified by AI and the considerations above would apply.

While this synergic relation between the two processes can only be proven with dedicated simulations in which CRs are active components (hybrid or MHD+PIC simulations), it is worth stressing that hybrid simulations have already shown evidence for episodes in which particle trapping in large magnetic structures occurs upstream, before these structures are eventually advected downstream \cite[]{SLAMS2025}. It is possible that such structures may be the result of overdense regions being unstable in the CR pressure gradient.


\section{Conclusions}\label{sec:concl}

CR-induced MFA is an integral part of the process of particle acceleration at shocks, especially in the case of SNR shocks. In this article we investigated the possibility that the amplification may be due to the reaction of small density perturbations to the CR pressure gradient in the upstream region of the shock, as originally suggested by DD. We expanded on previous investigations carried out using MHD simulations with an assigned pressure gradient (DD, VLS) by improving on resolution, on the discussions of numerical damping and the role of dimensionality, and by the adoption of parameters' values that are closer to the ones expected for a realistic SNR shock.

Our conclusions can be summarized as follows: 1) AI can lead to the nonlinear growth of small perturbations even for initial perturbations $\delta\rho/\rho_0\sim 10^{-1}$. This phenomenon leads to considerable MFA, especially when the original field is perpendicular to the shock normal. In contrast, previous works started with large initial perturbations, so that the role of AI was somehow reduced and nonlinear effects started immediately. We provided expressions for the growth rate and the scales involved. 2) The detailed study of the dependence of our results on resolution showed that by increasing the resolution, more power appears at large wavenumbers, as expected based on AI, but even the highest resolution runs were incapable of properly resolving the small scales where equipartition of kinetic and magnetic power is expected, for the realistic values of $\xi_{\rm CR}$ and Mach number $M_s$ adopted here. This consideration applies even more to previous investigations. As a consequence, we could only impose a lower limit to the magnitude of MFA. 3) When AI leads to $\delta \rho/\rho_0\gg 1$, we occasionally observe the formation of shocklets that temporarily stall and are later advected with the fluid across the shock. These structures look similar to the SLAMS known to the space plasma community \cite[]{SLAMS1992} and may play an important role for particle acceleration. 4) For values of the Mach number appropriate to a SNR, we showed that AI may grow faster than the non-resonant instability and lead to a larger MFA. We comment on the possibility that Bell instability may cooperate by first creating regions with $\delta\rho/\rho_0\gtrsim 1$ far upstream, and that these regions may induce further amplification due to the CR pressure gradient. Future investigation, possibly using an MHD+PIC approach, is planned to better study the nonlinear reaction of accelerated particles to this instability.


\begin{acknowledgements}
The authors are grateful to an anonymous referee for their comments that helped us improve the manuscript. They also acknowledge precious conversations with E.~Amato, V.~Berta, N.~Bucciantini, D.~Caprioli, B.~Olmi. They are also thankful to L. Drury and T.P. Downes for clarifications on their earlier work. The work of PB and AC was partially funded by the European Union - Next Generation EU under PRIN-MUR 2022TJW4EJ ``Unveiling the footprints of the cosmic ray journey through the Galaxy and beyond''. The work of PB was also partially funded under MUR National Innovation Ecosystem grant ECS00000041 - VITALITY/ASTRA - CUP D13C21000430001. The work of ES was supported by a Rita Levi Montalcini fellowship.
\end{acknowledgements}

\vfill\null
\bibliographystyle{aa}
\bibliography{refs}

\begin{appendix}

\section{Velocity perturbations for $k_y$ modes}\label{app:ky-velocity}

In this Appendix, we derive Eq.~(\ref{ky-withB}). We consider small density perturbations along $y$,
\begin{equation}\label{eq:ky-density-perturb}
    \rho(x,y)=\frac{\rho_0 +\delta\rho(x)\sin(k_y y)}{U(x)}~.
\end{equation}
We assume that the density perturbations are nearly independent of $x$, namely $\delta\rho\simeq\delta\rho_0$  (see Fig.~\ref{fig:ky}). The perturbed magnetic field is $\mathbf{B} = B_{y,0}U^{-1}\,\hat{\mathbf{y}} + \delta\mathbf{B}$, and the perturbed velocity is $\mathbf{u} = u_0U(x)\,\hat{\mathbf{x}} + \delta\mathbf{u}$.

Since $\xi_{\rm CR}\ll 1$, the derivatives of $U$ can be neglected with respect to the derivatives of the perturbations. Using $\nabla \cdot\delta\mathbf{B}=0$, the steady-state induction equation, $\nabla \times(\mathbf{u} \times\mathbf{B})=0$, can be approximated as
\begin{align}
    &\frac{\partial \delta B_y}{\partial x} = - \frac{B_{y,0}}{u_0 U^2}\frac{\partial \delta u_x}{\partial x}~,\label{eq:induction1}\\
    &\frac{\partial \delta B_x}{\partial x}=\frac{B_{y,0}}{u_0 U^2}\frac{\partial \delta u_x}{\partial y}~.\label{eq:induction2}
\end{align}
Assuming $v_{A,y}^2=B_{y,0}^2/4\pi\rho_0 \ll u_0^2$, and using Eq.~(\ref{eq:induction1}), the momentum equation can be approximated as
\begin{align}
    u_0U\frac{\partial \delta u_x}{\partial x} =  U\frac{\nabla P_{\rm CR}}{\rho_0}\frac{\delta\rho_0}{\rho_0}\sin(k_y y) + \frac{B_{y,0}}{4\pi\rho_0}\frac{\partial \delta B_x}{\partial y}~,\label{eq:momentum-appendix}
\end{align}
Making the ansatz $\delta u_x = a(x)\sin(k_y y)$, Eq.~(\ref{eq:induction2}) gives
\begin{equation}
    \frac{\partial \delta B_x}{\partial x}=a(x)\frac{k_yB_{y,0}}{u_0 U^2} \cos(k_y y)~.
\end{equation}
Taking the partial derivative of (\ref{eq:momentum-appendix}) with respect to $x$, we get
\begin{equation}
\label{eq:addt}
   a''(x) = -\frac{k_y^2B_{y,0}^2}{4\pi\rho_0 u_0^2U^3}\,a(x)~.
\end{equation}
Solving Eq.~\eqref{eq:addt} with BCs $a(0)=0$ and $a'(0)=(1/u_0)(\nabla P_{\rm CR}/\rho_0)(\delta\rho_0/\rho_0)$ yields
\begin{equation}\label{eq:final-deltaux}
    \delta u_x(x,y) = \frac{1}{u_0}\frac{\nabla P_{\rm CR}}{\rho_0}\frac{\delta\rho_0}{\rho_0}\frac{\sin(k_x'x)}{k_x'}\sin(k_yy)~,
\end{equation}
where $k_x'= k_y(v_A/u_0)U^{-3/2}$.

\section{Fourier transforms and power spectra}
\label{app:powerspec}

This Appendix outlines the formalism behind the discrete FT (DFT) and the power spectrum definitions used in Sect.~\ref{sec:sims}. Below, $f(\mathbf{x})$ denotes a generic field, representing density or a single velocity/magnetic field component. When specifically dealing with vector field components, we use Greek letter subscripts, $\mathbf{f}=(f_\alpha)$.

Our 2D simulation box has area $A=L^2/8$ embedded within a grid of $N_{\rm tot}=N_xN_y$ points with uniform spacings along each direction, $\Delta x=L/N_x$ and $\Delta y=L/8N_y$, and spatial coordinates
\begin{equation}\label{eq:grid-x}
    \mathbf{x}_{ij} = (x_i,y_j) = (i\Delta x,j\Delta y)~,\quad \begin{cases}
        i=0,...,N_x-1\\
        j=0,...,N_y-1 
    \end{cases}~.
\end{equation}
This grid allows for specific wavenumbers along each direction:
\begin{equation}
    \mathbf{k}_{ab} = (k_a,k_b) = \frac{2\pi}{L}(a,8b)~, \quad \begin{cases}
        a=-\frac{N_x}{2}+1,...,\frac{N_x}{2}\\
        b=-\frac{N_y}{2}+1,...,\frac{N_y}{2}
    \end{cases}~.
\end{equation}
Grid cell areas are $\Delta^2 x = \Delta x\Delta y$ and $\Delta^2 k = (2\pi)^2/A$ in physical and reciprocal spaces, respectively. Note that, along each direction, the maximum allowed $k$ corresponds to the Nyquist limit, $k_{x,max}=2\pi/2\Delta x$ and $k_{y,max}=2\pi/2\Delta y$, while the minimum allowed $k$ (besides the zero mode) corresponds to a wavelength the size of each box dimension, $k_{x,min}=2\pi/L$ and $k_{y,min}=2\pi/(L/8)=16\pi/L$.

We adopt the convention relating a two-dimensional field $f(\mathbf{x})$ to its FT $\tilde{f}(\mathbf{k})$ via
\begin{align}
    &\tilde{f}(\mathbf{k}) = \frac{1}{(2\pi)^2}\int f(\mathbf{x})\,e^{-i\mathbf{k}\cdot\mathbf{x}}\,d^2\mathbf{x}~,\\
    &f(\mathbf{x}) = \int \tilde{f}(\mathbf{k})\,e^{i\mathbf{k}\cdot\mathbf{x}}\,d^2\mathbf{k}~.
\end{align}
Upon discretization, this yields the DFT
\begin{equation}\label{eq:DFT}
    \tilde{f}_{ab} = \frac{1}{N_{\rm tot}}\sum_{ij} f_{ij}\,e^{-i\mathbf{k}_{ab}\cdot\mathbf{x}_{ij}}~,\qquad  f_{ij} = \sum_{ab}\tilde{f}_{ab}\,e^{i\mathbf{k}_{ab}\cdot\mathbf{x}_{ij}} ~,
\end{equation}
where one defines $f_{ij} \equiv f(\mathbf{x}_{ij})$ and $\tilde{f}_{ab} \equiv \Delta^2 k\,\tilde{f}(\mathbf{k}_{ab})$, obeying Parseval's identity
\begin{equation}
    \sum_{ij}|f_{ij}|^2 = N\sum_{ab} |\tilde{f}_{ab}|^2~.
\end{equation}
These DFTs can be efficiently calculated using NumPy's `\texttt{fft}' package \citep{harris2020array} under the `\texttt{forward}' normalization convention. 

Under statistical homogeneity, the power spectrum tensor $P_{\alpha\beta}(\mathbf{k}_{ab})$ is defined for vector fields via the ensemble average
\begin{equation}
    \Delta^2 k\,\langle \tilde{f}_\alpha(\mathbf{k}_{ab}) \tilde{f}_\beta^*(\mathbf{k}_{a'b'})\rangle = \delta_{aa'}\delta_{bb'}P_{\alpha\beta}(\mathbf{k}_{ab})~,
\end{equation}
or, equivalently, as the FT of the correlation tensor
\begin{equation}
    \xi_{\alpha\beta}(\mathbf{r}_{ij})=\langle f_\alpha(\mathbf{x})f^*_\beta (\mathbf{x}-\mathbf{r}_{ij})\rangle~.
\end{equation}
Scalar fields yield a scalar power spectrum, $P(\mathbf{k}_{ab})$, with an equivalent definition dropping the component indices. Rigorously performing ensemble averages requires many simulations, which is too computationally expensive. The best one can do from a single simulation is compute the estimator $\hat{P}_{\alpha\beta}(\mathbf{k}_{ab})=\Delta^2 k\,\tilde{f}_\alpha(\mathbf{k}_{ab}) \tilde{f}^*_\beta(\mathbf{k}_{ab})=\tilde{f}_{\alpha,ab}\tilde{f}^*_{\beta,ab}/\Delta^2 k$, which is the FT of $\hat{\xi}_{\alpha\beta}(\mathbf{r}_{ij}) = (1/N) \sum_{kl}  f_\alpha(\mathbf{x}_{kl})f^*_\beta (\mathbf{x}_{kl}-\mathbf{r}_{ij})$. Note that its trace at zero separation is
\begin{equation}
    \hat{\xi}_{\alpha\alpha}(\mathbf{0}) = \frac{1}{N}\sum_{ij} |\mathbf{f}(\mathbf{x}_{ij})|^2 = \sum_{ab} \Delta^2 k\,\hat{P}_{\alpha\alpha}(\mathbf{k}_{ab})~.
\end{equation}
We shall drop the hats from now on, always working with the estimators. For a statistically homogeneous and isotropic two-dimensional field satisfying $\nabla\cdot\mathbf{f}=0$, one can write \citep{Batchelor:1953fqo}
\begin{equation}
    P_{\alpha\beta}(\mathbf{k})=P(k)\left(\delta_{\alpha\beta} - \frac{k_\alpha k_\beta}{k^2}\right)~,
\end{equation}
where $P(k)=P_{\alpha\alpha}(\mathbf{k})=\Delta^2 k\,|\tilde{\mathbf{f}}(\mathbf{k})|^2$ is the tensor's trace, which depends only on $k=|\mathbf{k}|$. For a scalar field, isotropy also implies that $P(\mathbf{k}) = P(k)$. 

From $P(k_{ab})$ computed for either a scalar or a vector field $f$, one can define an omnidirectional power spectrum $E(k)$ (also commonly called an ``energy spectrum'') by binning $k$ and adding all the power within each bin $k_{ab}=|\mathbf{k}_{ab}|\in [k_n,k_{n+1} = k_n+\Delta k_n)$:
\begin{equation}\label{eq:energyspec}
    E(\bar{k}_n)\Delta k_n = C \sum_{\substack{ab \\ k_{ab}\in [k_n,k_{n+1})}} \Delta^2 k \,P(k_{ab})~,
\end{equation}
where $\bar{k}_n = \sqrt{k_n k_{n+1}}$ is a representative $k$ in bin $n$ (\textit{i.e.} the geometric mean of its edges) and $C$ is a constant prefactor depending on the specific field. This prefactor is introduced such that $\sum_n E(\bar{k}_n) \Delta k_n$ is an energy density, or some other quantity of interest, with $E(\bar{k}_n)\Delta k_n$ being the contribution coming from bandwidth $[k_n,k_n+\Delta k_n]$. In this work, we consider 4 types of omnidirectional power spectra:
\begin{enumerate}
    \item For density fluctuations, the field is $f=\delta\rho/\rho_0$, and we choose $C=1$ such that
    \begin{align}
        \sum_n &E_{\delta\rho/\rho_0}(\bar{k}_n) \Delta k_n = \sum_{ab}\Delta^2 k \, P(k_{ab}) = \xi(\mathbf{0}) \nonumber \\
        &= \frac{1}{A}\sum_{ij}\Delta x \Delta y\,\left[\frac{\delta\rho}{\rho_0}(\mathbf{x}_{ij})\right]^2 = \left(\frac{\delta\rho_{rms}}{\rho_0}\right)^2~,\label{eq:rhoPS}
    \end{align}
    where $\delta\rho_{rms}/\rho_0$ is the RMS $\delta\rho/\rho_0$ over all scales in the box. 
    \item For magnetic fields fluctuations, $\mathbf{f}=\delta\mathbf{B}$, we choose $C=1/8\pi$ whereby the turbulent magnetic energy density power spectrum $E_{\delta B}(k)\equiv E_{\rm turb,mag}(k)$ satisfies
    \begin{equation}
        \sum_n E_{\rm turb,mag}(\bar{k}_n) \Delta k_n = \frac{1}{A}\sum_{ij}\Delta x \Delta y\,\frac{|\delta\mathbf{B}(\mathbf{x}_{ij})|^2}{8\pi} = \frac{\delta B_{rms}^2}{8\pi}
    \end{equation}
    is the average magnetic energy density in perturbations at all scales, while $E_{\rm turb,mag}(\bar{k}_n) \Delta k_n = \delta B_{rms}^2(\bar{k}_n)/8\pi$ is that coming from perturbations with wavenumbers $k\in[k_n,k_n+\Delta k_n]$. In other words,
    \begin{equation}\label{eq:deltaBRMS}
        \delta B_{rms}(\bar{k}_n) = \sqrt{8\pi E_{\rm turb,mag}(\bar{k}_n) \Delta k_n}
    \end{equation}
    is the RMS value of magnetic field fluctuations with characteristic scale $\ell = 2\pi/\bar{k}_n$.
    \item For velocity fields, \textit{e.g.} $\delta\mathbf{u}$, we choose $C=\rho_0/2$ such that
    \begin{equation}
        \sum_n E_{\delta u}(\bar{k}_n) \Delta k_n = \frac{1}{A}\sum_{ij}\Delta x \Delta y\,\frac{\rho_0|\delta\mathbf{u}(\mathbf{x}_{ij})|^2}{2} = \frac{\rho_0 \delta u_{rms}^2}{2}~.
    \end{equation}
    Analogously to the magnetic field case, the RMS amplitude of velocity perturbations at a characteristic scale $\ell = 2\pi/\bar{k}_n$ can be found through
    \begin{equation}\label{eq:deltauRMS}
        \delta u_{rms}(\bar{k}_n) = \sqrt{\frac{2E_{\delta u}(\bar{k}_n) \Delta k_n}{\rho_0}}~.
    \end{equation}
    \item For studies of compressible turbulence, it is common to consider the density-weighted velocity $\mathbf{w}=\sqrt{\rho}\mathbf{u}$ to define kinetic energy spectra \citep{1990JSCom...5...85K,Grete:2017usl}. Let $\delta \mathbf{w}$ be its fluctuations around a mean background profile. We can then choose $C=1/2$ such that its corresponding power spectrum $E_{\delta w}(k) \equiv E_{\rm turb,kin}(k)$ is the turbulent kinetic energy spectrum:
    \begin{equation}
        \sum_n E_{\rm turb,kin}(\bar{k}_n) \Delta k_n = \frac{1}{A}\sum_{ij}\Delta x \Delta y\,\frac{|\mathbf{\delta w}(\mathbf{x}_{ij})|^2}{2}~.
    \end{equation}
\end{enumerate}
In practice, we calculate each $E(k)$ by defining 50 logarithmically spaced $k$-bins between $k_{min}=\max(k_{x,min},k_{y,min})$ and $k_{max}=\min(k_{x,max},k_{y,max})$, and add their power following Eq.~(\ref{eq:energyspec}). This leads to
\begin{equation}\label{eq:logbinning}
    E(k)\Delta k \approx kE(k)\Delta \ln k  = kE(k) \frac{\ln(k_{max}/k_{min})}{50}~,
\end{equation}
with $\ln(k_{max}/k_{min})/50 \approx 0.1$ in our simulations. This means that $\delta B_{rms}(k)$ and $\delta u_{rms}(k)$ can be estimated, within factors of order unity, from the logarithmic energy spectrum $kE(k)$ via:
\begin{align}\label{eq:deltaBk}
    \delta B_{rms}(k) & \sim \sqrt{kE_{\delta B}(k)}~,\\
\label{eq:deltauk}
    \delta u_{rms}(k) & \sim \sqrt{kE_{\delta u}(k)}~.
\end{align}

\subsection{Generating Initial Density Perturbations}
\label{app:density}

In Sects.~\ref{sec:manymodes} and \ref{sec:nonlinear}, we inject a density perturbations field $\delta\rho_0(\mathbf{x})/\rho_0$ which is superposition of modes with different wave vectors, according to Eq. (\ref{eq:DDdensity}), following a predetermined omnidirectional power spectrum (with $C=1$) between $\max(k_{x,min},k_{y,min})<k<\min(k_{x,max},k_{y,max})$,
\begin{equation}\label{eq:deltarhoPS}
    E_{\delta\rho/\rho_0}(k) = E^* \left(\frac{k}{k^*}\right)^\alpha
\end{equation}
and zero outside this range, where $E^*$ is a normalization setting the power in fluctuations at a reference scale $2\pi/k^*$. We fix $k^* = \min(k_{x,max},k_{y,max})$. Using the formalism presented above, we now describe our method for generating such a field\footnote{We actually generate an isotropic $\delta\rho_0/\rho_0$ field in 3D space with the desired power spectrum $E_{\rm 3D}\left(k=\sqrt{k_\perp^2 + k_z^2}\right) = E_0\, k^\alpha$, then take a 2D slice of it along the $xy$ plane to be the injected field into the box. It can be shown that the 2D slice's omnidirectional spectrum $E_{\rm 2D}(k_\perp)$ obeys the same power law as the 3D one as long as $\alpha<1$.}. This is done mode by mode in Fourier space, rather than directly in physical space, since it makes the connection with the desired power spectrum more direct.

We consider density perturbations to be Gaussian in nature, meaning that $\delta\tilde{\rho}_{ab}/\rho_0 =\Delta^2 k\,\delta\tilde{\rho}(\mathbf{k}_{ab})/\rho_0 = X_{ab} + i Y_{ab}$, with $X_{ab}$ and $Y_{ab}$ randomly sampled from a Gaussian distribution with zero mean and variance $\sigma^2_{ab}$. Writing this in polar form,
\begin{equation}\label{eq:fourierdeltarho}
    \frac{\delta \tilde{\rho}_{ab}}{\rho_0} = A_{ab}\,e^{i\phi_{ab}}~,
\end{equation}
it can be easily shown that $A_{ab}$ follows a Rayleigh distribution with scale parameter $\sigma_{ab}$, implying $A_{ab,rms}=\sqrt{2}\,\sigma_{ab}$, while $\phi_{ab}$ is uniformly distributed between $[0,2\pi)$. Recall that producing a real $\delta\rho(\mathbf{x}_{ij})/\rho_0$ requires $\delta\tilde{\rho}(-\mathbf{k}_{ab}) = \delta\tilde{\rho}^*(\mathbf{k}_{ab})$. This allows us to generate only half of the $k$-modes; once we have $\delta\tilde{\rho}_{ab}/\rho_0$ for all $a,b\geq 0$, the modes with $a,b<0$ are all set by the reality condition\footnote{Self-symmetric modes in which $a=-a$ mod $N_x$ and $b=-b$ mod $N_y$ must be real ($Y_{ab}=0$) and are therefore sampled directly from a Gaussian distribution with zero mean and $2\sigma^2_{ab}$ variance.}: $A_{-a,-b}=A_{ab}$ and $\phi_{-a,-b} = -\phi_{ab}$.

The power spectrum resulting from Eq.~(\ref{eq:fourierdeltarho}) is $P(k_{ab}) = A_{ab}^2/\Delta^2 k$, while its inverse FT yields
\begin{equation}
    \frac{\delta\rho}{\rho_0}(\mathbf{x}_{ij}) = \sum_{ab}A_{ab}\cos(\mathbf{k}_{ab}\cdot\mathbf{x}_{ij} + \phi_{ab})~,
\end{equation}
in agreement with Eq.~(\ref{eq:DDdensity}). The RMS density perturbation over all scales can be be shown to match our expectation:
\begin{equation}\label{eq:EstarfromRMS}
    \left(\frac{\delta\rho_{\rm rms}}{\rho_0}\right)^2 =\frac{1}{N}\sum_{ij}\left[\frac{\delta\rho}{\rho_0}(\mathbf{x}_{ij})\right]^2 = \sum_{ab}A_{ab}^2 = E^*\sum_n \left(\frac{\bar{k}_n}{k^*}\right)^\alpha\Delta k_n~.
\end{equation}
Crucially, Eq.~(\ref{eq:EstarfromRMS}) allows us to find $E^*$ for any desired value of $\delta\rho_{\rm rms}/\rho_0$. Meanwhile, the RMS value of perturbations on a scale $k_{ab}$ is simply $A_{ab,{\rm rms}}$.
The power inside each $k$-bin is
\begin{equation}
    E_{\delta\rho/\rho_0}(\bar{k}_n)\Delta k_n = \sum_{\substack{ab \\ k_{ab}\in [k_n,k_{n+1})}}A_{ab}^2 = N_n \overline{A_{ab}^2}~,
\end{equation}
where $N_n \simeq 2\pi \bar{k}_n\,\Delta k_n/\Delta^2 k$ is the number of grid points in the bin and $\overline{A_{ab}^2}$ is the bin-averaged power per mode, which we set to $A_{ab,{\rm rms}}$. In the $\Delta k_n\to 0$ limit, this approaches 
\begin{equation}
    A_{ab,{\rm rms}} = \sqrt{2}\sigma_{ab}=\sqrt{\frac{E_{\delta\rho/\rho_0}(k_{ab})}{2\pi k_{ab}}\Delta^2k}~.
\end{equation}
This relation determines $\sigma_{ab}$ used to randomly sample each $A_{ab}$ mode, guaranteeing that the resulting spectrum will be properly normalized and follow the desired scaling with $k$. Once we have sampled every $\delta\tilde{\rho}(\mathbf{k}_{ab})/\rho_0$, we simply take its inverse FT to get the final $\delta\rho(\mathbf{x}_{ij})/\rho_0$ entering the precursor.

\end{appendix}
\end{document}